\documentclass[11pt]{elsarticle}
\usepackage[margin=1in]{geometry}
\usepackage{amssymb,amsfonts,amsmath,mathrsfs,mathtools}
\usepackage{graphicx,epsfig,subfigure}
\usepackage{bm}
\usepackage{times}
\usepackage{multirow}
\usepackage{xcolor}
\usepackage{float}
\usepackage{algorithm}
\usepackage{algorithmicx}
\usepackage{algpseudocode}
\usepackage{listings,comment}
\usepackage{braket}
\usepackage[title]{appendix}
\usepackage{cleveref}

\newcommand{\rev}[1]{{#1}}

\newcommand\extrafootertext[1]{%
    \bgroup
    \renewcommand\thefootnote{\fnsymbol{footnote}}%
    \renewcommand\thempfootnote{\fnsymbol{mpfootnote}}%
    \footnotetext[0]{#1}%
    \egroup
}

\begin{document}

\title{Sampling two-dimensional isometric tensor network states}
\author[lbnl]{Alec Dektor\corref{cor1}}
\ead{adektor@lbl.gov}
\author[ornl]{Eugene Dumitrescu}
\ead{dumitrescuef@ornl.gov}
\author[lbnl]{Chao Yang}
\ead{cyang@lbl.gov}

\address[lbnl]{Applied Mathematics and Computational Research Division, Lawrence Berkeley National Laboratory, Berkeley, CA, USA}
\address[ornl]{Computational Sciences and Engineering Division, Oak Ridge National Laboratory, Oak Ridge, TN, USA}
\cortext[cor1]{Corresponding author}

\begin{abstract}
Sampling a quantum system's underlying probability distributions is an important computational task, e.g., for quantum advantage experiments and quantum Monte Carlo algorithms. Tensor networks are an invaluable tool for efficiently representing states of large quantum systems with limited entanglement. Algorithms for sampling one-dimensional (1D) tensor networks are well-established and utilized in several 1D tensor network methods. In this paper we introduce two novel sampling algorithms for two-dimensional (2D) isometric tensor network states (isoTNS) that \rev{generalize existing 1D tensor networks sampling algorithms}. \rev{Our first proposed} algorithm performs independent sampling and yields a single configuration together with its associated probability. The second algorithm employs a greedy search strategy to identify $K$ high-probability configurations and their corresponding probabilities. Numerical results demonstrate the effectiveness of these algorithms across quantum states with varying entanglement and system size. 
\end{abstract}

\maketitle

\extrafootertext{This manuscript has been authored by UT-Battelle, LLC, under Contract No.\ DE-AC0500OR22725 with the U.S.\ Department of Energy. The United States Government retains, and the publisher, by accepting the article for publication, acknowledges that the United States Government retains a nonexclusive, paid-up, irrevocable, worldwide license to publish or reproduce the published form of this manuscript, or allow others to do so, for the United States Government purposes. The Department of Energy will provide public access to these results of federally sponsored research in accordance with the DOE Public Access Plan.}

\section{Introduction} \label{sec:intro}
Sampling from probability distributions encoded by quantum systems has recently emerged as an important computational task, with both practical and wide-ranging theoretical implications. For example, from a complexity theory perspective, the existence of a classical algorithm that efficiently samples bitstrings from commuting instantaneous quantum polynomial (IQP complexity class) circuits in two-dimensions (2D) would imply the collapse of the polynomial hierarchy~\cite[Corollary 1]{Bremner2011} and fundamentally reshape our understanding of computation. In a practical vein, sampling bitstrings explicitly mimics Born's rule in quantum mechanics. This has applications in device engineering, including quantum digital twinning, and in computation, such as evaluating claims of experimental quantum supremacy~\cite{Arute2019} \rev{and generative modeling~\cite{Vieijra2022}}. In addition to these applications, sampling is itself an important classical subroutine. For example, sampling is used to model finite-temperature quantum systems in the minimally entangled typical thermal states (METTS)~\cite{White2009, Stoudenmire2010} algorithm. 

Tensor network algorithms are state-of-the-art linear algebraic tools pushing the limits of representing and simulating quantum systems. Algorithms for 1D quantum systems are, via the matrix product state (MPS) formalism~\cite{Schollwock_2011} and their canonical forms, at a mature stage. Sampling algorithms for MPS are well-developed and utilized in the study of 1D systems with tensor networks~\cite{Lami2023, Stoudenmire2010}. Related algorithms have also been proposed in the tensor train (TT) formalism~\cite{Dolgov2020,Chertkov2022,Batsheva2023,Novikov21}, a 1D tensor network that is mathematically equivalent to MPS,  for uncertainty quantification and generative modeling. However, the MPS is not well suited for systems with dimension $D \geq 2$. Several tensor network structures have been proposed to push the state of the art for simulation of quantum systems beyond 1D. Notable examples include multi-scale entanglement renormalization ansatze (MERA)~\cite{Ferris2012} and projected-entangled pair state (PEPS)~\cite{Verstraete2004}. 
\rev{Several algorithms sampling such 2D tensor networks have been proposed. For example, a direct PEPS sampling algorithm was proposed in \cite{Vieijra2021}, which was subsequently used for generative modeling of image data \cite{Vieijra2022}. A PEPS based trajectory sampling scheme for rare events in classical stochastic systems was proposed in \cite{Causer2023}, and a sampling using boundary MPS which can be applied to PEPS was proposed in \cite{Rudolph2025}. }

Recently, isometric tensor network states (isoTNS) have been proposed as a two-dimensional tensor network ansatz that resembles PEPS while incorporating isometric constraints analogous to the canonical forms of MPS~\cite{Haghshenas2019, Hyatt2020, zaletel2020isometric}. These constraints make the computation of norms, marginal distributions, local observables, and the application of local gates efficient once the orthogonality center is positioned appropriately within the network. In contrast to MPS, however, where the orthogonality center can be shifted by a single SVD, relocating the orthogonality center in a 2D isoTNS is substantially more involved. Variational and greedy methods such as the Moses Move~\cite{zaletel2020isometric}, based on multiple SVDs and unitary disentangler optimization~\cite{Wei2025}, inevitably introduce approximation errors. 
\rev{Nevertheless, isoTNS algorithms have attractive computational and algorithmic advantages compared to general PEPS. Their isometric structure reduces contractions on the orthogonality hypersurface to effectively one-dimensional problems and improves the conditioning of local updates, while standard PEPS expectation values rely on approximate boundary/environment contractions with a separate environment bond dimension in addition to the PEPS bond dimension. IsoTNS methods have been successfully applied to simulations of dynamics~\cite{Lin2022}, excited states~\cite{Dektor2025}, fermionic systems~\cite{dai2025fermionic, Wu2025}, thermal states~\cite{Kadow2023}, and string-net liquids~\cite{Soejima2020}. 
Variants such as diagonal isoTNS have also been proposed for non-square lattices~\cite{Sappler2025}.}

\rev{Previously considered direct PEPS and boundary-MPS sampling methods target arbitrary 2D tensor networks and rely on approximate environments or importance-sampling constructions. Sampling algorithms for MERA were shown to } 
In this paper, we introduce two efficient sampling algorithms for 2D isoTNS that \rev{exploit exact local marginals on the orthogonality hypersurface}. The first algorithm is a sampling algorithm that returns a single configuration and its associated probability. The second algorithm is a greedy search that determines $K$ high-probability configurations and their corresponding probabilities. Conceptually, both algorithms are generalizations of 1D tensor network algorithms to 2D isoTNS. The proposed algorithms share many desirable properties with their 1D counterparts, such as polynomial scaling in the virtual bond-dimensions. However, unlike MPS algorithms, the 2D isoTNS algorithms have an additional source of truncation error. We perform numerical experiments to quantify the effects of this truncation error in the context of sampling algorithms. \rev{We also note that a discussion of isoTNS sampling, with respect to computation complexity rather than generating bitstring samples, was presented in \cite{Malz2025}. }

The remainder of the paper is organized as follows. To build intuition for our isoTNS algorithms, Section~\ref{sec:background} briefly reviews tensor network notation and discusses algorithms, generating both independent and clusters of samples within the MPS formalism. In Section~\ref{sec:2d}, we propose algorithms, for independent samples and for obtaining several high-probability states, from a distribution represented by a two-dimensional isoTNS. In Section~\ref{sec:numerics}, we provide several numerical demonstrations which validate our proposed algorithms and analyze sampling error scaling. Our main findings and future directions are summarized in Section~\ref{sec:conclusions}. 

\section{Background: Tensor network methods} \label{sec:background}

Consider a quantum many-body system composed of $L$ spins. The wavefunction is generally written as
\begin{equation} \label{eq:state}
\ket{\Psi} =
\sum_{{\sigma_i}}
T(\sigma_1, \ldots, \sigma_L)
\ket{\sigma_1 \cdots \sigma_L},
\end{equation}
where $\ket{\sigma_1 \cdots \sigma_L}$ denotes a basis vector of a tensor-product Hilbert space. Each local index $\sigma_j$ takes d possible values 
, corresponding to a physical dimension $d$ per site. The tensor $T \in \mathbb{C}^{d \times \cdots \times d}$ contains the complex coefficients of the wavefunction $\ket{\Psi}$, written in the $\ket{\sigma_1 \cdots \sigma_L}$ basis. The number of coefficients grows exponentially in system size ($L^d$) which makes explicit storage or computation directly involving the coefficient tensor $T$ intractable. To alleviate this curse of dimensionality, tensor network methods are commonly used to represent low-rank states
In many situations, such as area-law entangled states, accurate tensor network representation can be reduced from exponential in $L$ to polynomial in $L$. In the worst case, the exponential scaling is recovered, e.g., for volume-law entangled states. 

Below, we examine efficient algorithms for sampling configurations from certain classes of 1D and 2D tensor networks. Concretely, sampling means generating a classical product state configuration $\sigma_1,\ldots,\sigma_L$ with probability $p(\sigma_1,\ldots,\sigma_L) = |T(\sigma_1,\ldots,\sigma_L)|^2$. We first build intuition by recalling sampling algorithms the 1D matrix product state (MPS) tensor network. In the subsequent section \ref{sec:2d}, this intuition is carried over and generalized to sampling algorithms for 2D isometric tensor network states isoTNS. 

\subsection{Matrix Product State (MPS)} 
\label{sec:1d}
Consider first an MPS, which is a tensor network with connectivity reflecting a one-dimensional lattice. A state \eqref{eq:state} is a MPS if its coefficients are factorized as 
\begin{equation} \label{eq:MPS}
\begin{aligned}
    T(\sigma_1,\ldots,\sigma_L) &= A_1^{\sigma_1} A_2^{\sigma_2} \cdots A_L^{\sigma_L} \\
    &=
    \raisebox{-0.74\height}{\includegraphics[width=0.3\textwidth]{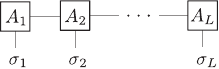}}
\end{aligned}
\end{equation}
where $A_i^{\sigma_i}$ is a matrix associated with the physical index $\sigma_i$. 
For $i = 2,\ldots,L-1$, the tensors $A_i^{\sigma_i}$ are $\chi \times \chi$ matrices, and the boundary tensors $A_1^{\sigma_1}$ and $A_L^{\sigma_L}$ are of size $1 \times \chi$ and $\chi \times 1$, respectively. The bond-dimension $\chi$ quantifies the amount of entanglement entropy in the state~\eqref{eq:state}. We use diagrammatic notation to represent tensor networks, where each tensor is depicted as a vertex (either a square or circle), and its indices (of a given dimension) are shown as edges (legs emanating from that shape). Connecting legs between tensors is a graphical Einstein summation notation and represents a contraction over the connected index shared by two tensors. To simplify diagrams in the rest of the paper, we often omit these physical index labels, as they can usually be inferred from the tensor labels. 

\subsubsection{MPS canonical form}
The tensors in the representation of Eq.~\eqref{eq:MPS} are not unique. MPS algorithms rely on canonical forms where a single tensor is chosen as the orthogonality center. 
The tensors in the MPS representation~\eqref{eq:MPS} can be transformed into canonical form with orthogonality center at site-$1$ by performing LQ decompositions on matrices obtained by reshaping the three-dimensional tensors $A_i$~\cite{Schollwock_2011} 
\begin{equation} \label{eq:mps_lq}
\begin{aligned}    {\includegraphics[width=0.4\textwidth]{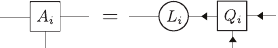}}. 
\end{aligned}
\end{equation}
Here the tensors $Q_i$ are isometries, meaning that multiplying $Q_i$ with its conjugate results in the identity. In this diagrammatic notation, isometries are indicated with arrows so that contracting a tensor with its conjugate over all indices with incoming arrows yields the identity operator in the space of all indices with outgoing arrows. Performing a sequence of LQ decompositions from right-to-left and absorbing the $L_i$ into neighboring tensors 
the MPS \eqref{eq:MPS} can always be brought into the canonical (or isometric) form 
\begin{equation} \label{eq:mps_can_1}
\begin{aligned}
T(\sigma_1,\ldots,\sigma_L) 
= \raisebox{-0.55\height}
{\includegraphics[width=0.31\textwidth]{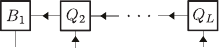}},
\end{aligned}
\end{equation}
where $B_1$ is referred to as the center of orthogonality. Due to the isometries, the \rev{global} norm of the MPS is given by the norm of $B_1$. As we shall demonstrate in the subsequent sections, canonical forms such as \eqref{eq:mps_can_1} \rev{provide a particularly transparent representation for computing marginal distributions (e.g., Eq.~\eqref{eq:mps_marg1}) used in sampling algorithms, although equivalent quantities can also be obtained from precomputed boundary contractions instead of leveraging canonical forms of the MPS.} 
The computational cost of transforming a MPS into isometric form with a sequence of QR-decompositions scales as $\mathcal{O}(dL\chi^3)$. For more details on MPS canonical form see~\cite{Perez-Garcia2007}. 

\subsubsection{Sampling from MPS} \label{sec:sampling_MPS}

To sample from the MPS \eqref{eq:MPS}, \rev{we use a canonical form representation \eqref{eq:mps_can_1}. Exact sampling can also be formulated directly in terms of precomputed boundary contractions, but the canonical form provides a particularly simple recursive construction of the required marginal and conditional distributions.} 
We assume the state \eqref{eq:mps_can_1} is normalized, otherwise it can be normalized by normalizing $B_1$. 
The probability of measuring $\sigma_1 \cdots \sigma_L$ from the state \eqref{eq:state} is given by Born's rule 
\begin{equation} \label{eq:born}
\begin{aligned}
p(\sigma_1,\ldots,\sigma_L) &= |T(\sigma_1,\ldots,\sigma_L)|^2 \\
&= \raisebox{-0.55\height}
{\includegraphics[width=0.3\textwidth]{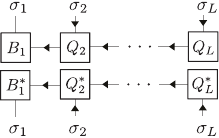}}, 
\end{aligned}
\end{equation}
where $\ast$ denotes element-wise complex conjugation. 
We now leverage the MPS structure in \eqref{eq:mps_can_1} to efficiently sample from the joint probability distribution \eqref{eq:born}. To do so, factor the distribution into a product of marginal and conditional distributions 
\begin{equation} \label{eq:prod_rule}
    p(\sigma_1,\sigma_2,\ldots,\sigma_L) = 
    p(\sigma_1) p(\sigma_2 \mid \sigma_1) \cdots p(\sigma_L \mid \sigma_1 \cdots \sigma_{L-1}), 
\end{equation}
and proceed by sampling each factor on the right-hand side of \eqref{eq:prod_rule}. 
The marginal distribution $p(\sigma_1)$ is obtained by tracing out $\sigma_2,\ldots,\sigma_L$ in \eqref{eq:born} which can be expressed as 
\begin{equation} \label{eq:mps_marg1}
\begin{aligned}
    p(\sigma_1) &= \raisebox{-0.55\height}{\includegraphics[width=0.3\textwidth]{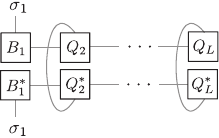}} \\
    &= \raisebox{-0.6\height}
{\includegraphics[width=0.1\textwidth]{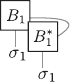}}
\end{aligned}
\end{equation}
where the second equality is due to the isometric property of the tensors $Q_2,\ldots,Q_L$, i.e., we have leveraged the MPS canonical form to obtain a simple representation of the marginal distribution $p(\sigma_1)$. 
We draw an independent sample $s_1$ from the marginal distribution \eqref{eq:mps_marg1} and store the corresponding probability $p_1=p(s_1)$. Then evaluate the MPS tensor $B_1$ at the sampled index $s_1$ by contracting the physical index with the $d \times 1$ standard basis vector $e_{s_1}$, and scale by a factor of $1/\sqrt{p_1}$ to obtain 
\begin{equation} \label{eq:evaluate_A1}
\begin{aligned}
{\includegraphics[width=0.25\textwidth]{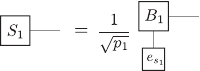}}. 
\end{aligned}
\end{equation}
The coefficient tensor \eqref{eq:mps_can_1} with $\sigma_1$ fixed at the selected index $s_1$ is represented by the tensor network 
\begin{equation} \label{eq:mps_left_can_s1}
\begin{aligned}
T(\sigma_2,\ldots,\sigma_L | s_1) &= \raisebox{-0.6\height}
{\includegraphics[width=0.3\textwidth]{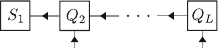}} 
\\
&= \raisebox{-0.55\height}
{\includegraphics[width=0.22\textwidth]{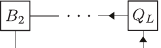}}, 
\end{aligned}
\end{equation}
where in the second line we defined 
\begin{equation} \label{eq:A2_def}
{\includegraphics[width=0.3\textwidth]{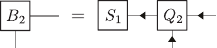}}. 
\end{equation}
Once again, by Born's rule, the conditional probability $p(\sigma_2, \ldots, \sigma_L \mid s_1)$ is the squared modulus of \eqref{eq:mps_left_can_s1} and the marginal $p(\sigma_2 \mid s_1)$ appearing as the second factor on the right-hand side of \eqref{eq:prod_rule} is obtained by tracing $\sigma_3,\ldots,\sigma_L$ out of this conditional probability. Thanks to the tensors $Q_3,\ldots,Q_L$ being isometries, such marginal is easily computed from $B_2$ 
\begin{equation} \label{eq:p2_mps}
\begin{aligned}
p(\sigma_2 \mid s_1) 
= \raisebox{-0.6\height}
{\includegraphics[width=0.1\textwidth]{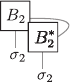}}
\end{aligned}
\end{equation}
We draw an independent sample $s_2$, from the marginal distribution \eqref{eq:p2_mps}, and store the corresponding probability $p_2=p(\sigma_2 \mid s_1)$. Then evaluate the MPS tensor $B_2$ at the sampled index $s_2$, by contracting the physical index with the $d \times 1$ standard basis vector $e_{s_2}$, and scale by a factor of $1/\sqrt{p_2}$ to obtain $S_2$ analogous to \eqref{eq:evaluate_A1}. Finally, contracting $S_2$ with $Q_{3}$ yields $B_3$, which can then be used to sample an index from the third site of the MPS. 

Continuing this process recursively, we draw a single independent sample from each of the marginal distributions on the right-hand side of \eqref{eq:prod_rule} and hence obtain one independent sample from the joint distribution on the left. This yields a perfect sampling algorithm \cite{Djuric2002}. 
\rev{In the canonical form implementation adopted here, the cost of first transforming the MPS into the form \eqref{eq:mps_can_1} and then sequentially sampling all sites scales as $\mathcal{O}(dL\chi^3)$. We emphasize, however, that canonization is not strictly necessary for exact MPS sampling. Other perfect sampling schemes can be formulated using precomputed boundary contractions with equivalent asymptotic cost but may offer practical constant factor speedups. We adopt the canonical form viewpoint because it generalizes naturally to isoTNS, where isometric structure is built into the ansatz and allows analogous simplifications of local marginals on the orthogonality hypersurface. The main steps of the canonical form version are summarized in Algorithm~\ref{alg:mps_sample}. }

Note that the algorithm described above performs sampling in the computational (canonical) basis. It is straightforward to modify this algorithm to sample in an arbitrary basis. For example, instead of sampling the first marginal distribution \eqref{eq:mps_marg1}, which corresponds to the diagonal entries of the reduced density matrix 
$$
\rho(\sigma_1, \sigma_1') = B_1^{\sigma_1} B_1^{\sigma_1'},
$$
a change of basis may be performed by applying a \rev{$d \times d$} unitary matrix $U$, yielding the transformed reduced density matrix
$$
\rho_U(\gamma_1,\gamma_1') = (U B_1)^{\gamma_1} (B_1^{\ast} U^{\ast})^{\gamma_1'},
$$
with corresponding sampling probabilities
$$
p_U = \operatorname{diag}(\rho_U).
$$
One natural choice for $U$ is the unitary that diagonalizes the reduced density matrix $\rho$. Similarly, basis transformations can be performed before sampling any of the subsequent marginal distributions $B_j$ for $j=2,\ldots,L$.

\begin{algorithm}[H]
\caption{Sampling from a MPS.}
\label{alg:mps_sample}
\begin{algorithmic}[1]
\Require 
\Statex $T = B_1 Q_2 \cdots Q_L$ $\rightarrow$ normalized MPS in canonical form 
\Ensure 
\Statex $s = s_1 s_2 \cdots s_L$ $\rightarrow$ independent sample from the distribution $|T|^2$ 
\Statex $p_L$ $\rightarrow$ probability associated with sample $s$
\For{$i=1$ to $L$}
    \State $s_i$ $\gets$ independent sample from $p(\sigma_i \mid s_1,\ldots,s_{i-1}) = \operatorname{Tr} \left[ B_i^{\sigma_i} \left(B_i^{\sigma_i}\right)^{\ast}\right]$
    \State ${p}_i$ $\gets$ $p(s_i \mid s_1 \cdots s_{i-1})$
    \If{$i < L$}
        \State $S_i$ $\gets$ project out $s_i$ from $B_i$ and scale by $1/\sqrt{p_i}$ 
        \State $B_{i+1}$ $\gets$ contract $S_i$ with $Q_{i+1}$
    \EndIf
\EndFor
\end{algorithmic}
\end{algorithm}

\subsubsection{MPS greedy search for $K$ high-probability configurations}
\label{sec:top_K_MPS}

The MPS sampling Algorithm~\ref{alg:mps_sample} described in the previous section returns a single sample $s$ drawn from the distribution \eqref{eq:born}, together with its associated probability. In some settings, e.g., in Monte Carlo tensor network algorithms such as METTS~\cite{White2009, Stoudenmire2010}, a single independent sample is sufficient. In other contexts, however, it is desirable to explore the distribution in a more structured manner. 

While Algorithm~\ref{alg:mps_sample} can be applied repeatedly to empirically estimate the global probability distribution, this approach can be inefficient. A more efficient algorithm is obtained by modifying the MPS sampling procedure to perform a greedy search for $K$ high-probability configurations of the distribution \eqref{eq:born}. An algorithm of this type was originally proposed in~\cite{Chertkov2022} for tensor train (TT), \rev{in \cite{Jalowiecki2021} for MPS, and in \cite{Rams2021} for generic PEPS. 
We emphasize that these methods and the top-$K$ search methods presented hereafter do not guarantee the highest probabilities. 
An algorithm that guarantees the highest probabilities would require backtracking to test all branches with high enough marginal probabilities, and in the worst case, such algorithm is exponential in $L$.} 

Rather than drawing a single random sample $s_i$ from each marginal distribution in Algorithm~\ref{alg:mps_sample}, we select the $K$ most probable outcomes at every step and append them to the high-probability partial bitstrings constructed in previous steps. The goal is for each site tensor to be projected onto these $K$ selected partial configurations. This is achieved by introducing an additional $K$-dimensional index that labels the retained partial bitstrings and is propagated through the MPS as the algorithm proceeds. The associated probabilities of the partial bitstrings are tracked concurrently. We now present the detailed steps of the resulting top-$K$ MPS algorithm. 

We begin with a normalized MPS in canonical form \eqref{eq:mps_can_1}. Next, construct the first marginal distribution \eqref{eq:mps_marg1} as before. Instead of drawing a random sample $s_1$ as we have done before, we select $K$ indices $\bm s_1$ of \eqref{eq:mps_marg1} with the highest probabilities $\bm p_1=p(\bm s_1)$. Then we evaluate $B_1$ at the $K$ high-probability indices by contracting with the $d \times K$ matrix $E_{\bm s_1}$ whose columns are standard basis vectors corresponding to the high-probability indices 
\begin{equation} \label{eq:evaluate_A1_K}
\begin{aligned}
{\includegraphics[width=0.25\textwidth]{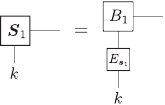}}
\end{aligned}
\end{equation}
The coefficient tensor \eqref{eq:mps_can_1} with the site-1 indices $\sigma_1$ constrained to the selected top-$K$ indices $\bm s_1$ is represented by the tensor network  
\begin{equation} \label{eq:mps_left_can_s1_K}
\begin{aligned}
T(k, \sigma_2,\ldots,\sigma_L) &= \raisebox{-0.7\height}
{\includegraphics[width=0.3\textwidth]{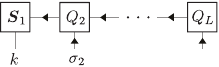}} 
\\
&= \raisebox{-0.55\height}
{\includegraphics[width=0.22\textwidth]{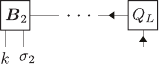}}
\end{aligned}
\end{equation}
where in the second line we defined 
\begin{equation} \label{eq:A2_def_K}
{\includegraphics[width=0.3\textwidth]{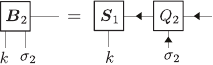}} 
\end{equation}
Note that the tensor $\bm B_{2}$ has two physical indices, $k$ corresponds to the $K$ high-probability partial bitstrings $\bm s_1$ obtained from sampling the first physical index and $\sigma_2$ is the physical index of site $2$. 
Now we can obtain the probabilities of the partial bitstrings $\sigma_{1}\sigma_{2}$ where $\sigma_{1}$ is restricted to the $K$ bitstrings $\bm s_1$ selected in the previous step of the algorithm 
\begin{equation} \label{eq:p2_mps_K}
\begin{aligned}
p(k, \sigma_2) 
= \raisebox{-0.6\height}
{\includegraphics[width=0.1\textwidth]{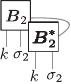}}. 
\end{aligned}
\end{equation}
To obtain $K$ high-probability bitstrings of length two from $\bm s_1$ and \eqref{eq:p2_mps_K}, we select the $K$ multi-indices $\{(k^{(m)},\sigma_2^{(m)})\}_{m=1}^K$ with the largest probabilities and construct the length-2 strings by appending $\sigma_2^{(m)}$ to the $k^{(m)}$-th element of $\bm s_1$
$$
\bm s_2(m) = \bm s_1\left(k^{(m)}\right) \sigma_2^{(m)}, \qquad m = 1,2,\ldots,K. 
$$
The corresponding probability of each bitstring is 
$$
\bm p_2(m) = \bm p_1\left(k^{(m)}\right) p\left(k^{(m)}, \sigma_2^{(m)}\right). 
$$
Then we evaluate $\bm B_2$ at the $K$ high-probability multi-indices by contracting with the tensor $E_{\bm s_2}$ obtained from reshaping the $K d \times K$ matrix whose columns are standard basis vectors corresponding to the selected multi-indices
\begin{equation} \label{eq:evaluate_A2_K}
\begin{aligned}
{\includegraphics[width=0.25\textwidth]{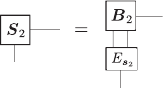}}. 
\end{aligned}
\end{equation}
To setup the tensors for constructing length 3 high-probability bitstrings, we contract $\bm S_{2}$ with $Q_{3}$ to obtain $\bm B_{3}$. 

By proceeding recursively, from left-to-right across the MPS, we obtain a list $\bm s_L$ of $K$ high-probability length-$L$ bitstrings and the corresponding probabilities $\bm p_L$. The main steps of the MPS top-$K$ algorithm are summarized in Algorithm~\ref{alg:mps_top_K}. Just as in the MPS sampling algorithm, we have presented the top-$K$ algorithm with probabilities selected in the canonical basis. It is straightforward to modify the algorithm to obtain high-probability strings other different bases by performing a local change of basis before selecting high-probability partial strings at each step. 

\begin{algorithm}[H]
\caption{Greedy search for $K$ high-probability configurations of MPS.}
\label{alg:mps_top_K}
\begin{algorithmic}[1]
\Require 
\Statex
$T = B_1 Q_2 \cdots Q_L$ $\rightarrow$ normalized MPS in canonical form
\Statex 
$K$ $\rightarrow$ Desired number of high-probability states
\Ensure 
\Statex ${\bm s}_L$ $\rightarrow$ list of top-$K$ bitstrings of length $L$ 
\Statex $\bm p_L$ $\rightarrow$ probabilities associated with bitstrings $\bm s_L$
 \State $\bm s_1$ $\gets$ top-$K$ configurations from $p(\sigma_1) = \operatorname{Tr} \left[ B_1^{\sigma_1} \left(B_1^{\sigma_1}\right)^{\ast}\right]$
 \State ${\bm p}_1$ $\gets$ $p(\bm s_1)$
 \State $\bm S_1$ $\gets$ project out $\bm s_1$ from $B_1$
 \State $\bm B_{2}$ $\gets$ contract $\bm S_1$ with $Q_{i+1}$
\For{$i=2$ to $L$}
    \State $\left(k^{(m)}, \sigma_i^{(m)}\right)$ $\gets$ top-$K$ multi-indices of $p(k, \sigma_i) = \operatorname{Tr} \left[\bm B_i^{k,\sigma_i} \left(\bm B_i^{k,\sigma_i}\right)^{\ast}\right]$
    \State $\bm s_i(m)$ $\gets$ $\bm s_{i-1}\left(k^{(m)}\right) \sigma_i^{(m)}$
    \State $\bm p_i(m)$ $\gets$ $\bm p_{i-1}\left(k^{(m)}\right) p\left(k^{(m)}, \sigma_i^{(m)}\right)$ 
    \If{$i < L$}
        \State $\bm S_i$ $\gets$ project out $\bm s_i$ from $\bm B_i$
        \State $\bm B_{i+1}$ $\gets$ contract $\bm S_i$ with $Q_{i+1}$
    \EndIf
\EndFor
\end{algorithmic}
\end{algorithm}

\section{Two-dimensional tensor network states}
\label{sec:2d}

Now we consider states defined on a two-dimensional square lattice 
\begin{equation} \label{eq:state_2d}
\ket{\Psi} =
\sum_{{\sigma_{ij}}}
T(\sigma_{11}, \ldots, \sigma_{LL})
\ket{\sigma_{11} \cdots \sigma_{LL}}, 
\end{equation}
and correlation coefficients represented as a tensor network with connectivity reflecting the two-dimensional lattice. A possible form of such tensor network is the projected entangled pair state (PEPS)~\cite{Verstraete2004, Orus2014} where the coefficient tensor in \eqref{eq:state_2d} is expressed as 
\begin{equation} \label{eq:peps}
\begin{aligned}
T_{\mathrm{peps}}(\sigma_{11},\ldots,\sigma_{LL}) 
= \raisebox{-0.45\height}{\includegraphics[width=0.4\textwidth]{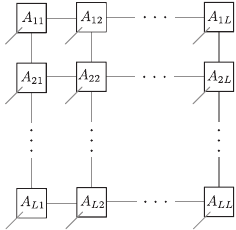}}. 
\end{aligned}
\end{equation}
Here each $A_j$ is, in general, a tensor with 5 legs and 4 neighbors, and edges connecting tensors indicate contractions (traces) over the corresponding virtual indices in the tensor network diagram. For simplicity, we have left off the physical index labels $\sigma_{ij}$ on the right-hand side, as they can be inferred by the tensor label $A_{ij}$. 

We consider a subset of two-dimensional tensor network states on a square lattice referred to as 2D isometric tensor network state (isoTNS)~\cite{zaletel2020isometric}
\begin{equation} \label{eq:2d_iso_tns}
\begin{aligned}
T(\sigma_{11},\ldots,\sigma_{LL}) 
= \raisebox{-0.45\height}{\includegraphics[width=0.4\textwidth]{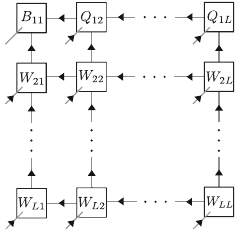}}. 
\end{aligned}
\end{equation}
The overall tensor network structure resembles PEPS. However, as indicated by the arrows, all tensors except one, in this case the top left tensor $B_{11}$, are isometries. The advantage of isoTNS \eqref{eq:2d_iso_tns} over PEPS \eqref{eq:2d_iso_tns}, is that contracting large parts of the tensor network with its dual is analytic and trivial. Hence, objects such as norms, marginal probability distributions, local observables, and applications of local gates, can be efficiently computed. The cost of many isoTNS algorithms, e.g., for moving the orthogonality center, scale as $\chi^7$. 

Intuitively, the isoTNS \eqref{eq:2d_iso_tns} can be thought of as a generalization of MPS canonical \eqref{eq:mps_can_1} form to two-dimensions. However, unlike with MPS, where the orthogonality center can be moved with a single SVD, moving the orthogonality center of a 2D isoTNS is more involved. Efficient algorithms for moving the orthogonality center, such as the greedy Moses Move~\cite{zaletel2020isometric}, involves two SVDs and the optimization of a unitary disentangling tensor~\cite{Wei2025}.  These extra steps introduce additional errors into the tensor network factorization. 
\rev{However, the advantage of enforcing isometric constraints is that contractions required, e.g., for computing norms, local observables, or marginals become extremely efficient. }

\subsection{Sampling from 2D isoTNS}
\label{sec:iso_tns_sample}
We now present a new algorithm for drawing an independent sample from the probability distribution defined by the modulus squared of the two-dimensional complex amplitudes in the isoTNS form \eqref{eq:2d_iso_tns} 
\begin{equation} \label{eq:born_2d}
\begin{aligned}
p(\sigma_{11},\ldots,\sigma_{LL}) &= |T(\sigma_{11},\ldots,\sigma_{LL})|^2.
\end{aligned}
\end{equation}
We assume that the isoTNS \eqref{eq:2d_iso_tns} has unit two-norm so that \eqref{eq:born_2d} is indeed a probability distribution. If not, the isoTNS can be normalized by normalizing $B_{11}$. For simplicity and without loss of generality, the algorithm presented hereafter assumes the center of orthogonality is in the top-left corner. However, the algorithm can be easily adapted to other isoTNS representations with orthogonality centers in other locations. 

Analogous to the 1D MPS case \eqref{eq:prod_rule}, we factor the joint distribution into a product of marginal and conditional distributions to leverage the isometric tensor network structure for efficient sampling 
\begin{equation} \label{eq:prod_rule_2d}
p(\sigma_{11}, \sigma_{12}, \ldots, \sigma_{LL}) = 
p(\sigma_{11}) p(\sigma_{12} \mid \sigma_{11}) \cdots p(\sigma_{LL} \mid \sigma_{11} \cdots \sigma_{(L-1)L})
\end{equation}
Here the column-major ordering in the sequence of conditional distributions is necessary for the algorithm presented below, which sweeps left to right across each \textit{row} of the 2D tensor network. If the orthogonality center begins in a different location of the 2D network or if one wishes to sweep up or down columns of the network, a different ordering of the variables in the product rule \eqref{eq:prod_rule_2d} may be used. This is analogous to the factorization \eqref{eq:prod_rule} for 1D networks, which lends itself to sweeping from left to right in the MPS. In this case, we start with an MPS in left-canonical form. If sweeping from right to left, one would begin with an MPS in right-canonical form. 


Generalizing the MPS algorithm, the idea is to leverage the isoTNS structure to efficiently construct the factors on the right-hand side of \eqref{eq:prod_rule_2d} recursively from left to right. We begin from \eqref{eq:2d_iso_tns} with the orthogonality center in the top-left corner, and start by sampling the first row 
\begin{equation} \label{eq:R1}
    R_1(\sigma_{11},\ldots,\sigma_{1L})
    = \raisebox{-0.6\height}
    {\includegraphics[width=0.3\textwidth]{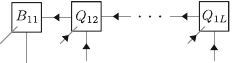}}.
\end{equation}
The steps used to sample the first row are then applied recursively to sample all rows of the isoTNS. 

The first marginal distribution $p(\sigma_{11})$ is obtained by tracing the remaining $\sigma_{ij}$ out of $TT^{\ast}$. For a general PEPS \eqref{eq:peps}, performing the tensor contractions required for such a trace is computationally demanding and, in general, must be approximated. However, thanks to the isometries in the isoTNS \eqref{eq:2d_iso_tns} these contractions are trivial,  and the marginal distribution is obtained exactly from just the $(1,1)$ site tensor 
\begin{equation} \label{eq:p11}
\begin{aligned}
p(\sigma_{11}) &= \text{Tr}_{(i,j\neq1,1)}[T(\sigma_{11},\ldots, \sigma_{LL})T^{\ast}(\sigma_{11},\ldots, \sigma_{LL})] \\
&= \raisebox{-0.6\height}
{\includegraphics[width=0.18\textwidth]{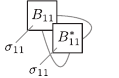}}. 
\end{aligned}
\end{equation}
Following the MPS sampling algorithm, we draw an independent sample $s_{11}$ from the marginal distribution \eqref{eq:p11} and store the corresponding probability $p_{11}=p(s_{11})$. As before, we then evaluate the tensor $B_{11}$ at the sampled index $s_{11}$ by contracting the physical index of $B_{11}$ with the $d \times 1$ standard basis vector $e_{s_{11}}$ and scale by a factor of $1/\sqrt{p_{11}}$ to obtain 
\begin{equation} \label{eq:evaluate_A11}
\begin{aligned}
{\includegraphics[width=0.3\textwidth]{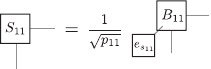}}. 
\end{aligned}
\end{equation}
The isoTNS row \eqref{eq:R1}, with the $\sigma_{11}$ index fixed at the selected index $s_{11}$, is now represented by the tensor network 
\begin{equation} \label{eq:2d_tns_1sample}
\begin{aligned}
R_1(\sigma_{12}, \ldots,\sigma_{1L} \mid s_{11}) 
= \raisebox{-0.45\height}{\includegraphics[width=0.3\textwidth]{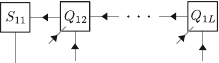}}
\end{aligned}
\end{equation}
Next we must construct and sample $p(\sigma_{12} \mid s_{11})$. If we were to follow the same procedure as we have done for MPS, we would contract $S_{11}$ with $Q_{12}$. However, that would result in a tensor with two vertical bonds, and proceeding in this way across the entire row yields a tensor with $L$ vertical bonds. As in the case of PEPS algorithms, such a procedure constructs a tensor with $L$ legs and thus scales exponentially with lattice side length $L$. Instead, to obtain an isoTNS row sampling algorithm that scales linearly with $L$, we move the center of orthogonality from the $(1,1)$ tensor to the $(1,2)$ tensor by performing a QR-decomposition on the tensor $S_{11}$ with the vertical bond considered as ``rows'' and the horizontal bond as ``columns'' 
\begin{equation}
{\includegraphics[width=0.3\textwidth]{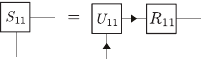}}
\end{equation}
and then absorbing $R_{11}$ into $Q_{12}$ to obtain 
\begin{equation}
{\includegraphics[width=0.3\textwidth]{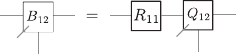}}. 
\end{equation}
The new $(1,2)$ site tensor $B_{12}$ is now the orthogonality center of the partially sampled isoTNS, and in particular it is the orthogonality center of the partially sampled first row 
\begin{equation} \label{eq:2d_tns_1sample_qr}
\begin{aligned}
R_1(\sigma_{12}, \ldots,\sigma_{1L} \mid s_{11}) 
= \raisebox{-0.45\height}{\includegraphics[width=0.3\textwidth]{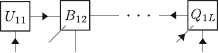}}. 
\end{aligned}
\end{equation}
The remaining tensors in the first isoTNS row are processed recursively from left to right in exactly the same way as the $(1,1)$ tensor to obtain and sample $p(\sigma_{1j} \mid s_{11} \cdots s_{1(j-1)})$ for $j=2,\ldots,L$. 

Let us explicitly describe the steps for $j=2$. Since the orthogonality center is at site $(1,2)$, we easily obtain the probability distribution $p(\sigma_{12} \mid s_{11})$ from $B_{12}$ 
\begin{equation} \label{eq:p21}
\begin{aligned}
p(\sigma_{12} \mid s_{11}) 
= \raisebox{-0.6\height}
{\includegraphics[width=0.15\textwidth]{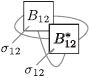}}. 
\end{aligned}
\end{equation}
Then sample \eqref{eq:p21} to obtain $s_{12}$ and the associated probability $p_{12} = p(s_{12} \mid s_{11})$. Next evaluate $B_{12}$ at $s_{12}$ by contracting it with the $s_{12}$ standard basis vector $e_{s_{12}}$ and normalize by $1/\sqrt{p_{12}}$ to obtain $S_{12}$, analogous to \eqref{eq:evaluate_A11} for the $(1,1)$ tensor. To prepare for the computation of the next distribution $p(\sigma_{13} \mid s_{11} s_{12})$, perform a QR decomposition of $S_{12}$ with the vertical bond considered as ``rows'' and the horizontal bond as ``columns'' to obtain $U_{12}$ and $R_{12}$. Then absorb $R_{12}$ into $Q_{13}$ to obtain $B_{13}$ as the new orthogonality center of the partially sampled isoTNS row 
\begin{equation} \label{eq:R1_B13}
\begin{aligned}
R_1(\sigma_{13}, \ldots,\sigma_{1L} \mid s_{11} s_{12}) 
= \raisebox{-0.45\height}{\includegraphics[width=0.35\textwidth]{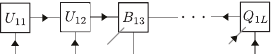}}. 
\end{aligned}
\end{equation}
Repeating this process for $j=3,\ldots,L$ yields the partial configuration $s_{11} \cdots s_{1L}$ and its probability $p_{1L}$. The main steps for sampling an isoTNS row are provided in Algorithm~\ref{alg:iso_row_sample}. 

Once the $(1,L)$ site tensor is sampled, the resulting isoTNS row has the form 
\begin{equation} \label{eq:2d_tns_Lsample}
\begin{aligned}
R_1(\mid s_{11},\ldots,s_{1L}) 
= \raisebox{-0.45\height}{\includegraphics[width=0.3\textwidth]{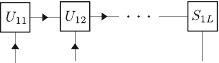}}
\end{aligned}
\end{equation}
with no physical indices remaining and orthogonality center on the right-hand side of the row at site $(1,L)$. 
In order to proceed to sampling the second isoTNS row, we contract the sampled first row \eqref{eq:2d_tns_Lsample} with the second row in \eqref{eq:2d_iso_tns} and ensure that the orthogonality center is placed on the left tensor of resulting row
\begin{equation} \label{eq:R2}
\begin{aligned}
R_2(\sigma_{21}, \ldots, \sigma_{2L} \mid s_{11} \cdots s_{1L}) &= \raisebox{-0.45\height}{\includegraphics[width=0.3\textwidth]{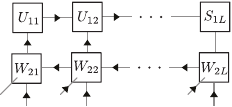}} \\
&\approx \raisebox{-0.45\height}{\includegraphics[width=0.3\textwidth]{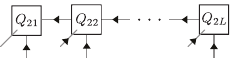}}.
\end{aligned}
\end{equation}
In general, directly contracting the two rows in the first line of \eqref{eq:R2} yields a new row with horizontal bond-dimension equal to the product of horizontal bond-dimensions from each row. To control the bond-dimension of the new row, this contraction can be performed approximately with controlled bond-dimension $\chi$ using any MPO-MPS multiplication algorithm, e.g., the zip-up algorithm~\cite{Stoudenmire2010}. In Section~\ref{sec:numerics} we examine the errors resulting from this MPO-MPS contraction. 

The second row \eqref{eq:R2} is sampled with the same steps we used on the first row \eqref{eq:R1}. We summarize the steps of sampling an isoTNS row in Algorithm~\ref{alg:iso_row_sample}. Proceeding recursively sweeping over all rows of the isoTNS we obtain a single sample $s_{11} \cdots s_{LL}$ of the joint probability distribution on the left-hand side of \eqref{eq:prod_rule_2d} and the associated probability $p_{LL}$. Note that once the final row is reached no vertical bonds remain. Thus the final row can be considered as an MPS and sampled using the method described in Section~\ref{sec:1d} and Algorithm~\ref{alg:mps_sample}. 

To summarize, we introduced a subroutine for efficient sampling of an isoTNS row for which the main steps are provided in Algorithm~\ref{alg:iso_row_sample}. After sampling each row, MPO-MPS multiplication is performed to prepare the subsequent row for sampling. Once the final row is reached, it can be viewed as an MPS and sampled using the MPS Algorithm~\ref{alg:mps_sample}. The main steps of the proposed isoTNS sampling algorithm are given in Algorithm~\ref{alg:iso_sample}. In this algorithm, $\text{sample row}(\bullet)$ is a call to Algorithm~\ref{alg:iso_row_sample} and $\text{MPO-MPS}[\bullet, \bullet, \chi]$ is an approximate MPO-MPS contraction routine with maximum bond-dimension $\chi$ that returns the approximate contraction and the Frobenius norm error. 

Just as in the MPS sampling algorithm presented in Section~\ref{sec:sampling_MPS}, we have presented the isoTNS sampling algorithm
with probabilities selected in the canonical basis. It is straightforward to modify the algorithm to sample in other different bases by performing a local change of basis before sampling each marginal distribution. 

\begin{algorithm}[htb]
\caption{Sampling from isoTNS row.}
\label{alg:iso_row_sample}
\begin{algorithmic}[1]
\Require 
\Statex $R_i = B_{i1} Q_{i2} \cdots Q_{iL}$ $\rightarrow$ row of partially sampled isoTNS with orthogonality center at $B_{i1}$. 
\Ensure 
\Statex $s_{i1}\cdots s_{iL}$ $\rightarrow$ a sample from the isoTNS row 
\Statex $p_{iL}$ $\rightarrow$ probability associated with sample 
\Statex $R_i(\mid s_{i1}\cdots s_{iL}) = U_{i1} \cdots U_{i(L-1)} S_{iL}$ $\rightarrow$ isoTNS row with samples projected out (no physical indices)
\For{$j = 1$ to $L$}
    \State $s_{ij}$ $\gets$ independent sample from $p(\sigma_{ij} \mid s_{11} \cdots s_{(i-1)j}) = \operatorname{Tr} \left[ B_{ij}^{\sigma_{ij}} \left(B_{ij}^{\sigma_{ij}}\right)^{\ast}\right]$
    \State $p_{ij}$ $\gets$ $p(s_{ij} \mid s_{11} \cdots s_{(i-1)j})$
    \If{$\rev{j} < L$}
        \State $S_{ij}$ $\gets$ project out $s_{ij}$ from $B_{ij}$ and scale by $1/\sqrt{p_{ij}}$ 
        \State $U_{ij}R_{ij}$ $\gets$ $\text{QR}(S_{ij})$
        \State $B_{i(j+1)}$ $\gets$ contract $R_{ij}$ with $Q_{i(j+1)}$
    \EndIf
\EndFor
\end{algorithmic}
\end{algorithm}

\begin{algorithm}[htb]
\caption{Sampling from isoTNS. }
\label{alg:iso_sample}
\begin{algorithmic}[1]
\Require 
\Statex $T$ $\rightarrow$ isoTNS \eqref{eq:2d_iso_tns} with orthogonality center at top left tensor. 
\Statex $\chi$ $\rightarrow$ maximum horizontal bond-dimension 
\Ensure 
\Statex $s_{11}\cdots s_{LL}$ $\rightarrow$ a sample from the isoTNS 
\Statex $p_{LL}$ $\rightarrow$ probability associated with sample 
\Statex $(e_1,\ldots,e_{L-1})$ $\rightarrow$ errors from contracting rows
\State $R_1$ $\gets$ $(B_{11}, Q_{12}, \ldots, Q_{1L})$
\For{$i = 1$ to $L$}
    \State $s_{i:}, p_{iL}, R_i(\mid s_{i1} \cdots s_{iL})$ $\gets$ $\text{sample row}(R_i)$
    \If{$i < L$}
        \State $R_{i+1}$, $e_i$ $\gets$ $\text{MPO-MPS}[R_i(\mid s_{i1} \cdots s_{iL}), W_{(i+1):}, \chi]$
    \EndIf
\EndFor
\end{algorithmic}
\end{algorithm}

\subsubsection{Computational cost}
\label{sec:cost_sample} 
Given an isoTNS with its orthogonality center in the top left corner, all physical dimensions equal to $d$, and all bond-dimensions at most $\chi$, we bound the computational cost of drawing a single sample using Algorithm~\ref{alg:iso_sample}. 

First, we estimate the cost of sampling a single row using Algorithm~\ref{alg:iso_row_sample}. In step 2 the marginal distribution $p(\sigma_{ij} \mid s_{11}, \ldots, s_{(i-1)j})$ is obtained by contracting a tensor containing at most one physical leg and three virtual legs (two horizontal and one vertical) with its adjoint (shown in \eqref{eq:p21} for $i,j=1,2$). Such marginal is computed by multiplying matrices with dimensions $d \times \chi^3$ and $\chi^3 \times d$ for a cost of $\mathcal{O}(d^2 \chi^3)$ operations. Projecting out a physical index in step 5 is computed by matrix vector product between the $s_{ij}$ standard basis vector of dimension $d \times 1$ and a matrix of size $d \times \chi^3$ (shown in \eqref{eq:evaluate_A11} for $i,j=1,1$) for a cost of $\mathcal{O}(d \chi^3)$. This cost is negligible compared to forming the marginal in step 2. In step 6 a QR decomposition is performed on $S_{ij}$ considered as a matrix with dimension at most $\chi^2 \times \chi$ for a cost of $\mathcal{O}(\chi^4)$. Finally, in step 7 we contract $R_{ij}$ with dimension $\chi \times \chi$ with $Q_{i(j+1)}$ considered as a matrix with dimension at most $\chi \times d \chi^2$. Such a contraction costs $\mathcal{O}(d\chi^4)$ operations. We perform $\mathcal{O}(L)$ of these steps and thus the total cost of sampling one isoTNS \rev{row} is $\mathcal{O}(L(d^2 \chi^3 + d\chi^4))$. 

Next we estimate the cost of the isoTNS sampling Algorithm~\ref{alg:iso_sample}. In step 3 we apply the row sampling algorithm, costing $\mathcal{O}(L(d^2 \chi^3 + d\chi^4))$ as described above. In step 5 we perform MPO-MPS multiplication where the length of the MPO/MPS is $L$, the dimension of each internal index is at most $\chi$, and the dimension of each external index is at most $d\chi$. There are different algorithms one can use to perform this multiplication. We use the zip-up algorithm~\cite{Stoudenmire2010} for a total cost of $\mathcal{O}(Ld^2\chi^6)$, which is the last step and dominates the cost of sampling a single row. Since we perform $L-1$ of these MPO-MPS multiplications, the total cost of the isoTNS sampling algorithm scales as $\mathcal{O}(L^2d^2\chi^6)$. Such computational complexity is lower than other standard isoTNS algorithms, such as the sequential Moses Move~\cite{zaletel2020isometric} for shifting the orthogonality center, which scales as $d^2\chi^7$. 

\subsubsection{2D isoTNS greedy search for $K$ high-probability configurations}
\label{sec:topK}

Next, by modifying the isoTNS sampling procedure in direct analogy with the modifications that lead from the MPS sampling algorithm to its top-$K$ variant discussed in Section~\ref{sec:top_K_MPS}, we develop an isoTNS top-$K$ algorithm. Starting from a normalized isoTNS with orthogonality center in the top-left corner \eqref{eq:2d_iso_tns}, we describe the procedure for obtaining top-$K$ strings of the first isoTNS row \eqref{eq:R1}. This procedure is then applied recursively to obtain $K$ high-probability configurations of the entire isoTNS \eqref{eq:2d_iso_tns}. 

First, construct the marginal density \eqref{eq:p11}. Then, instead of drawing one sample $s_{11}$ from this distribution as in the sampling algorithm, select $K$ indices $\bm s_{11}$ with the highest probabilities $\bm p_{11} = p(\bm s_{11})$. Then evaluate the tensor $B_{11}$ at the selected indices by contracting the physical index with the $d \times K$ matrix $E_{\bm s_{11}}$ whose columns are standard basis vectors corresponding to the high-probability indices 
\begin{equation} \label{eq:evaluate_A11_K}
\begin{aligned}
{\includegraphics[width=0.3\textwidth]{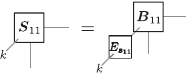}}
\end{aligned}
\end{equation}
Next we perform a QR decomposition on $\bm S_{11}$ with the vertical bond considered as ``rows'' and the horizontal bond and $k$ index as ``columns'' 
\begin{equation}
{\includegraphics[width=0.3\textwidth]{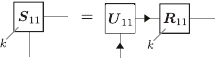}}
\end{equation}
Then contract $\bm R_{11}$ with $Q_{12}$ to move the orthogonality center and the sampled $k$ index to the $(1,2)$ site tensor 
\begin{equation}
{\includegraphics[width=0.3\textwidth]{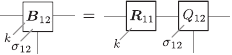}}
\end{equation}
The resulting tensor $\bm B_{12}$ has two physical indices. One labeled $k$, which indexes the $K$ high-probability partial bitstrings $\bm s_{11}$ obtained from sampling the $(1,1)$ physical index $\sigma_{11}$, and $\sigma_{12}$ which corresponds to the physical index of site $(1,2)$. Next we compute probabilities of the length-2 strings $\sigma_{11}\sigma_{12}$ with $\sigma_{11}$ restricted to the $K$ strings $\bm s_{11}$ selected in the previous step of the algorithm 
\begin{equation} \label{eq:p21_K}
\begin{aligned}
p(k, \sigma_{12}) 
= \raisebox{-0.6\height}
{\includegraphics[width=0.15\textwidth]{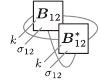}}. 
\end{aligned}
\end{equation}
To obtain $K$ high-probability bitstrings of length $2$ from \eqref{eq:p21_K}, we select $K$ multi-indices $\{(k^{(m)}, \sigma_{12}^{(m)})\}_{m=1}^K$ with largest probabilities and construct $K$ length-2 strings by appending $\sigma_{12}^{(m)}$ to $\bm s_{11}(k^{(m)})$ 
$$
\bm s_{12}^{(m)} = \bm s_{11}\left(k^{(m)}\right) \sigma_{12}^{(m)}, \qquad m = 1,2,\ldots,K. 
$$
The probability of the $m$-th string is given by 
$$
\bm p_{12}(m) = \bm p_{11}\left(k^{(m)}\right) p\left(k^{(m)}, \sigma_{12}^{(m)}\right). 
$$
Then we evaluate $\bm B_{12}$ at the $K$ high-probability multi-indices by contracting with the tensor $E_{\bm s_{12}}$ obtained from reshaping the $K d \times K$ matrix whose columns are standard basis vectors corresponding to the selected multi-indices 
\begin{equation} 
\begin{aligned}
{\includegraphics[width=0.3\textwidth]{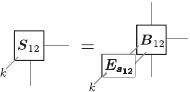}}
\end{aligned}
\end{equation}
where the index $k$ has dimension $K$ and now indexes length-2 high-probability strings in the list $\bm s_{12}$. 

The algorithm proceeds recursively across the first isoTNS row until we obtain $\bm s_{1L}$ containing $K$ high-probability strings of length $L$, the associated probabilities $\bm p_{1L}$ and an isoTNS row of the form 
\begin{equation} \label{eq:2d_tns_Lsample_K}
\begin{aligned}
{\includegraphics[width=0.3\textwidth]{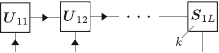}}
\end{aligned}
\end{equation}

In order to process the second isoTNS row, we contract the sampled first row \eqref{eq:2d_tns_Lsample_K} with the second row in \eqref{eq:2d_iso_tns} and ensure that the orthogonality center and $k$ index is placed on the left tensor of the resulting row 
\begin{equation} \label{eq:R2_K}
\begin{aligned}
\bm R_2(k, \sigma_{21}, \ldots, \sigma_{2L} \mid \bm s_{1L}) &= \raisebox{-0.45\height}
{\includegraphics[width=0.3\textwidth]{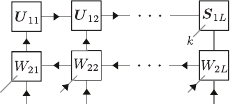}} \\
&\approx \raisebox{-0.45\height}{\includegraphics[width=0.3\textwidth]{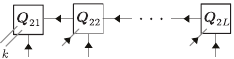}}.
\end{aligned}
\end{equation}
Just like the isoTNS sampling algorithm presented in Section~\ref{sec:iso_tns_sample}, directly contracting the two rows in the first line of \eqref{eq:R2_K} yields a new row with horizontal bond-dimension equal to the product of horizontal bond-dimensions from each row. To control the bond-dimension of the new row, this contraction can be performed approximately with controlled bond-dimension $\chi$ using any MPO-MPS multiplication algorithm, e.g., the zip-up algorithm~\cite{Stoudenmire2010}. In Section~\ref{sec:numerics} we examine the errors resulting from this MPO-MPS contraction. 

To obtain $K$ high probability strings of length $2L$, the second row \eqref{eq:R2_K} is processed with the same steps we used on the first row \eqref{eq:R1}. We summarize the steps of obtaining $K$ high-probability strings from an isoTNS row in Algorithm~\ref{alg:iso_row_topK}. In steps 2,3 with $j=0$ we let $i0 = (i-1)L$ in order to make the 2D indexing of the lattice compatible with the linear indexing of strings. Recursively sampling isoTNS rows from top to bottom, we obtain a list $\bm s_{LL}$ containing $K$ strings of length $L^2$ and their probabilities $\bm p_{LL}$. Note that no vertical bonds remain once the final row is reached . Thus the final row can be considered as an MPS and sampled using the top-$K$ MPS algorithm described in Section~\ref{sec:1d} and Algorithm~\ref{alg:mps_top_K}. 

\subsubsection{Computational cost}
The steps for the top-$K$ isoTNS algorithm are similar to the steps in the sampling algorithm presented in Section~\ref{sec:iso_tns_sample}, thus the computational complexity can be analyzed similarly. The main difference in the top-$K$ algorithm is that an additional index, of dimension at most $K$, is carried around on the tensor in the orthogonality center. 

First we analyze the computational cost of top-$K$ on an isoTNS row as described in Algorithm~\ref{alg:iso_row_topK}. \rev{In the following cost estimate, the candidate label $k$ is treated as an external branch index rather than absorbed into an enlarged virtual bond. In step 2, only the $dK$ diagonal entries of the reduced density matrix are required. These can be obtained as squared norms of the corresponding tensor slices, each containing at most one physical leg, one $k$ index, two horizontal legs, and a single vertical leg, for a cost of $\mathcal{O}(dK\chi^3)$, without forming the full $dK\times dK$ reduced density matrix. In step 5, selecting the retained multi-indices is a gather operation on the corresponding slices and costs $\mathcal{O}(K\chi^3)$. In step 6, the resulting $K$ tensors can be QR decomposed independently; each QR acts on a matrix of dimension at most $\chi^2 \times \chi$ and therefore costs $\mathcal{O}(\chi^4)$, for a total of $\mathcal{O}(K\chi^4)$. Finally, in step 7, the resulting $R$ tensors are contracted with $\bm Q_{i(j+1)}$ for a cost of $\mathcal{O}(dK\chi^4)$. We perform $L-1$ such steps and thus the total cost of sampling one isoTNS row is $\mathcal{O}\!\left(LdK\chi^4\right)$. }


Next, we estimate the cost of the entire isoTNS greedy top-$K$ algorithm~\ref{alg:iso_topK}. In step 3, we apply the row sampling algorithm \rev{above at cost $\mathcal{O}\!\left(LK(d\chi^3 + \chi^4 + d\chi^4)\right)$. In step 5, the MPO--MPS multiplication can likewise be organized independently for each of the $K$ retained candidates, rather than by fusing the candidate label into an enlarged bond dimension. Using the zip-up algorithm~\cite{Stoudenmire2010}, each multiplication costs $\mathcal{O}(Ld^2\chi^6)$, so performing it for all $K$ candidates costs $\mathcal{O}(LKd^2\chi^6)$. We perform $L-1$ of these MPO--MPS multiplications, and hence the total cost of the greedy top-$K$ isoTNS sampling algorithm scales as $\mathcal{O}(L^2 K d^2 \chi^6)$ which dominates the cost of row sampling.}


\begin{algorithm}[htb]
\caption{$K$ high-probability configurations from an isoTNS row.}
\label{alg:iso_row_topK}
\begin{algorithmic}[1]
\Require 
\Statex $\bm R_i(k,\sigma_{i1},\ldots,\sigma_{iL} \mid \bm s_{(i-1)L}) = \bm B_{i1} \bm Q_{i2} \cdots \bm Q_{iL}$ $\rightarrow$ row of partially sampled isoTNS with orthogonality center and $k$-index at $\bm B_{i1}$. 
\Statex $\bm s_{(i-1)L}$ $\rightarrow$ $K$ strings of length $(i-1)L$ 
\Statex $\bm p_{(i-1) L}$ $\rightarrow$ probabilities of each string 
\Ensure 
\Statex $\bm s_{iL}$ $\rightarrow$ $K$ strings of length $iL$
\Statex $\bm p_{iL}$ $\rightarrow$ probabilities of each string 
\Statex $\bm R_i(k \mid \bm s_{iL}) = \bm U_{i1} \cdots \bm U_{i(L-1)} \bm S_{iL}$ $\rightarrow$ sampled isoTNS row with $k$ index and orthogonality center at $\bm S_{iL}$
\For{$j = 1$ to $L$}
    \State $\left(k^{(m)}, \sigma_{ij}^{(m)}\right)$ $\gets$ top-$K$ multi-indices from 
    $p(k, \sigma_{ij} \mid \bm s_{i(j-1)}) = \operatorname{Tr} \left[\bm B_{ij}^{k,\sigma_{ij}} \left(\bm B_{ij}^{k,\sigma_{ij}}\right)^{\ast}\right]$
    \State $\bm p_{ij}(m)$ $\gets$ $\bm p_{i(j-1)}\left(k^{(m)}\right) p\left(k^{(m)}, \sigma_{ij}^{(m)} \mid \bm s_{i(j-1)}\right)$
    \If{$\rev{j} < L$}
        \State $\bm S_{ij}$ $\gets$ project out $\bm s_{ij}$ from $\bm B_{ij}$
        \State $\bm U_{ij} \bm R_{ij}$ $\gets$ $\text{QR}(\bm S_{ij})$
        \State $\bm B_{i(j+1)}$ $\gets$ contract $\bm R_{ij}$ with $\bm Q_{i(j+1)}$
    \EndIf
\EndFor
\end{algorithmic}
\end{algorithm}

\begin{algorithm}[htb]
\caption{$K$ high-probability configurations from isoTNS. }
\label{alg:iso_topK}
\begin{algorithmic}[1]
\Require 
\Statex $T$ $\rightarrow$ isoTNS \eqref{eq:2d_iso_tns} with orthogonality center at top left tensor 
\Statex $\chi$ $\rightarrow$ maximum horizontal bond-dimension 
\Ensure 
\Statex $\bm s_{LL}$ $\rightarrow$ list of $K$ high-probability strings
\Statex $\bm p_{LL}$ $\rightarrow$ probabilities of strings
\Statex $(e_1,\ldots,e_{L-1})$ $\rightarrow$ errors from contracting rows
\State $\bm R_1$ $\gets$ $(B_{11}, Q_{12}, \ldots, Q_{1L})$
\For{$i = 1$ to $L$}
    \State $\bm s_{iL}, \bm p_{iL}, \bm R_i(k \mid \bm s_{iL})$ $\gets$ $\text{top-K row}[\bm R_{i}(k, \sigma_{i1}, \ldots, \sigma_{iL} \mid \bm s_{(i-1)L})]$ 
    \If{$i < L$}
        \State $\bm R_{i+1}(k, \sigma_{(i+1)1}, \ldots, \sigma_{(i+1)L} \mid \bm s_{iL})$, $e_i$ $\gets$ $\text{MPO-MPS}[\bm R_i(k \mid \bm s_{iL}), W_{(i+1):}, \chi]$
    \EndIf
\EndFor
\end{algorithmic}
\end{algorithm}

\subsection{Errors in isoTNS sampling algorithms}

In both the independent isoTNS sampling algorithm (Algorithm~\ref{alg:iso_sample}) and the top-K isoTNS algorithm (Algorithm~\ref{alg:iso_topK}), the only source of approximation error arises from contracting a sampled row with its neighbor using MPO–MPS contraction methods (e.g., the zip-up algorithm). 

This occurs in step 5 of both algorithms. The first row is sampled before any row contractions are performed and is therefore sampled exactly. If all subsequent row contractions were performed without truncation error (i.e., $\epsilon_i = 0$ for $i = 1,\ldots,L-1)$, the entire isoTNS sampling procedure is exact. In practice, however, each MPO–MPS contraction can introduce a nonzero truncation error, which we denote by $\epsilon_i$. As a result, when sampling from the second column, we no longer work with the exact wavefunction conditioned on the previously sampled indices, but instead with an approximation whose Frobenius-norm error is determined by the truncation error incurred during the MPO–MPS contraction. The same situation arises for subsequent rows, with approximation errors accumulating as additional contractions are performed. 

In general, the sampling procedure operates on $T_{\chi}$, an approximation of the ideal correlation tensor $T$. Assuming the cumulative truncation error is bounded in the Frobenius norm as $\|T_{\chi} - T\|_F \leq \epsilon$, we may assess the induced error in the associated probability distribution $p_{\chi} = |T_{\chi}|^2$. Writing $T_{\chi} = T + \epsilon A$, for some perturbation tensor with $\|A\|_F = \mathcal{O}(1)$, we have $p_{\chi} = |T|^2 + \mathcal{O}(\epsilon)$. That is, for small $\epsilon$, truncation errors in the wavefunction propagate linearly into errors in the sampled probability distribution. 

\begin{figure}[t]
\centering
\includegraphics[scale=0.4]{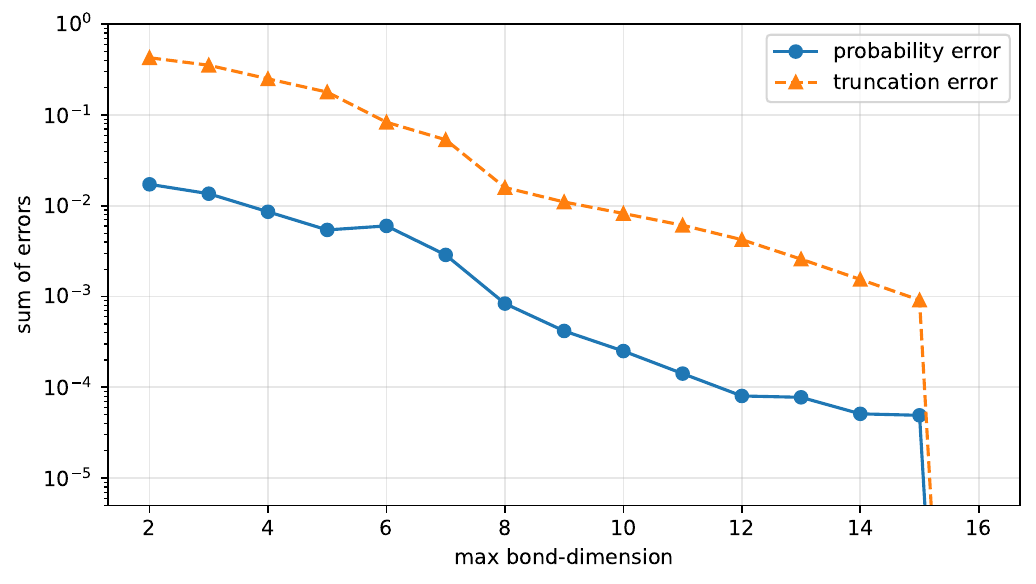}
\caption{top-$K$ isoTNS algorithm applied to random state. With isoTNS bond-dimension $1,2,\ldots,16$. We plot the sum of the column truncation errors and the sum of the $L1$ errors in the $K$ probabilities found by the isoTNS algorithm. } 
\label{fig:topK_rand_err_vs_bond_dim}
\end{figure}

While the behavior of MPO–MPS truncation errors in algorithms such as zip-up is well understood heuristically, they are not, in general, rigorously controlled. For this reason, we perform a series of numerical experiments to empirically assess the accuracy of the isoTNS sampling algorithms. In the following section, the results are presented in Figures~\ref{fig:rand_error}, \ref{fig:topK_rand}, \ref{fig:topK_rand_err_vs_bond_dim} and discussed in detail. 

\section{Numerical demonstrations} \label{sec:numerics}

We now present numerical demonstrations of the proposed isoTNS sampling algorithms. All algorithms were implemented using the quimb library~\cite{Gray2018} and the implementations are publicly available~\cite{isoTNSCode}. 

\subsection{Test states and their isoTNS constructions} 
To validate our 2D isoTNS sampling algorithm, we considered several states of qubits (spin $1/2$ particles) on a $L\times L$ lattice. 

As a special instance of ferromagnetic ordering, the GHZ-state 
\begin{equation} \label{eq:GHZ}
\ket{\mathrm{GHZ}} = \frac{1}{\sqrt{2}} \left( \ket{0}^{\otimes L^2} + \ket{1}^{\otimes L^2} \right),
\end{equation}
exhibits long-range correlations across the entire lattice, such that all qubits are aligned in either in the $\ket{0}$ or $\ket{1}$ state simultaneously. 
The W-state 
\begin{equation} \label{eq:W}
\ket{\mathrm{W}} = \frac{1}{L} \sum_{i=1}^{L^2} \ket{0}^{\otimes (i-1)} \ket{1}_i \ket{0}^{\otimes (L^2 - i)},
\end{equation}
represents a uniform superposition of all basis states with exactly one excitation. \rev{As isoTNSs, the GHZ-state \eqref{eq:GHZ} and W-state \eqref{eq:W} can be exactly represented with maximum bond-dimension $2$, independent of the lattice size. The construction of such isoTNS are described in \ref{sec:isoTNS_construction}. 
While encoding relatively simple probability distributions, i.e. that of a balanced coin and $L^2$ sided die respectively, these states' exact isoTNS representations on large lattices provides a preliminary, controlled, and scalable baseline test of the sampling algorithm implementation. 
} 

We also considered a state generated from a sequence of random unitary 2-site nearest neighbor gates constructed from the $SU(4)$ decomposition techniques described in~\cite{Khaneja2001}. Beginning from the vacuum state, we applied random nearest neighbor gates across each column starting from the left most column. Then we applied a sequence of random gates across each row from left to right starting from the top row. After applying all gates, we transformed the resulting state into an isoTNS with orthogonality center at the top left site using the sequential Moses Move. A sufficiently large bond-dimension was used in the Moses Move to not introduce any error while preparing the state as an isoTNS. 
\rev{Such randomly generated states are thus exactly represented as isoTNS, however due to their large bond-dimension, allow us to assess truncation errors made during the sampling algorithms. }

\subsection{Empirical distribution from independent samples}
\label{sec:empirical_results}
\begin{figure}[t]
\centering {\footnotesize \hspace{1.8cm} $8 \times 8$ lattice \hspace{5.7cm} $16 \times 16$ lattice } \\
\begin{minipage}{0.05\textwidth}
\centering
\rotatebox{90}{\footnotesize GHZ-state}
\end{minipage}
\begin{minipage}{0.45\textwidth}
\centering
\includegraphics[scale=0.4]{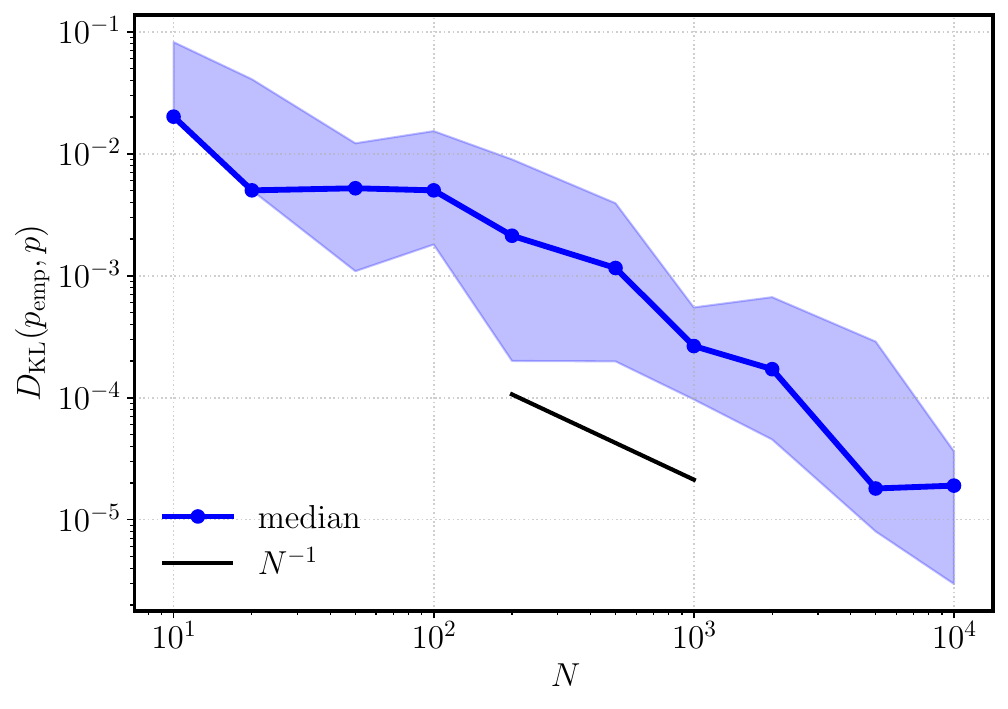}
\end{minipage}
\begin{minipage}{0.45\textwidth}
\centering
\includegraphics[scale=0.4]{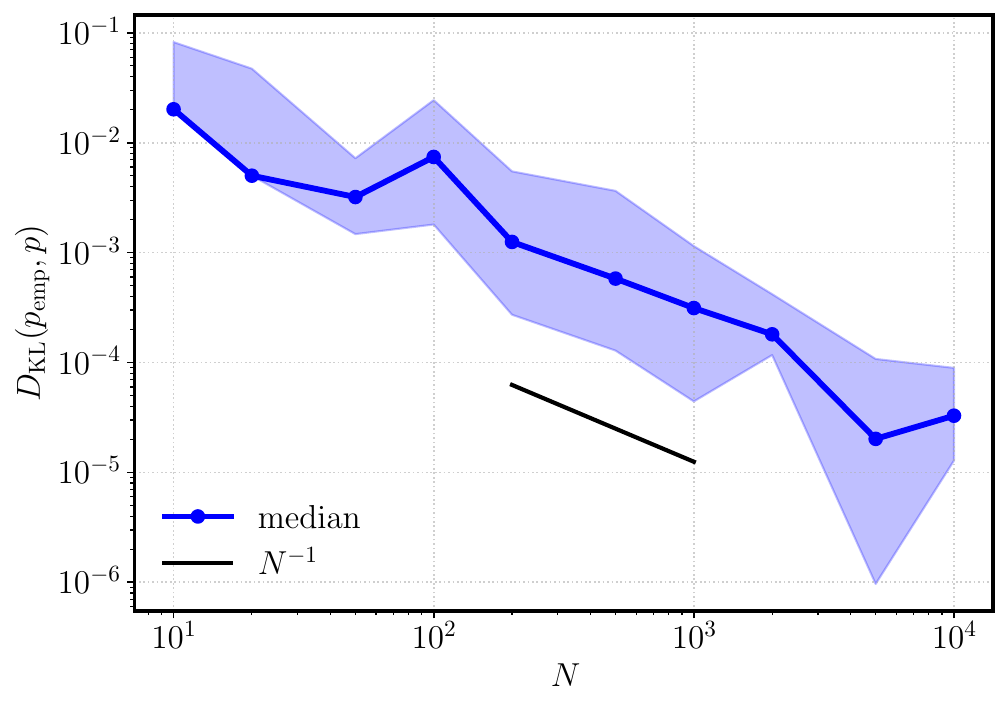}
\end{minipage}

\vspace{0.3cm}

\begin{minipage}{0.05\textwidth}
\centering
\rotatebox{90}{\footnotesize W-state}
\end{minipage}
\begin{minipage}{0.45\textwidth}
\centering
\includegraphics[scale=0.4]{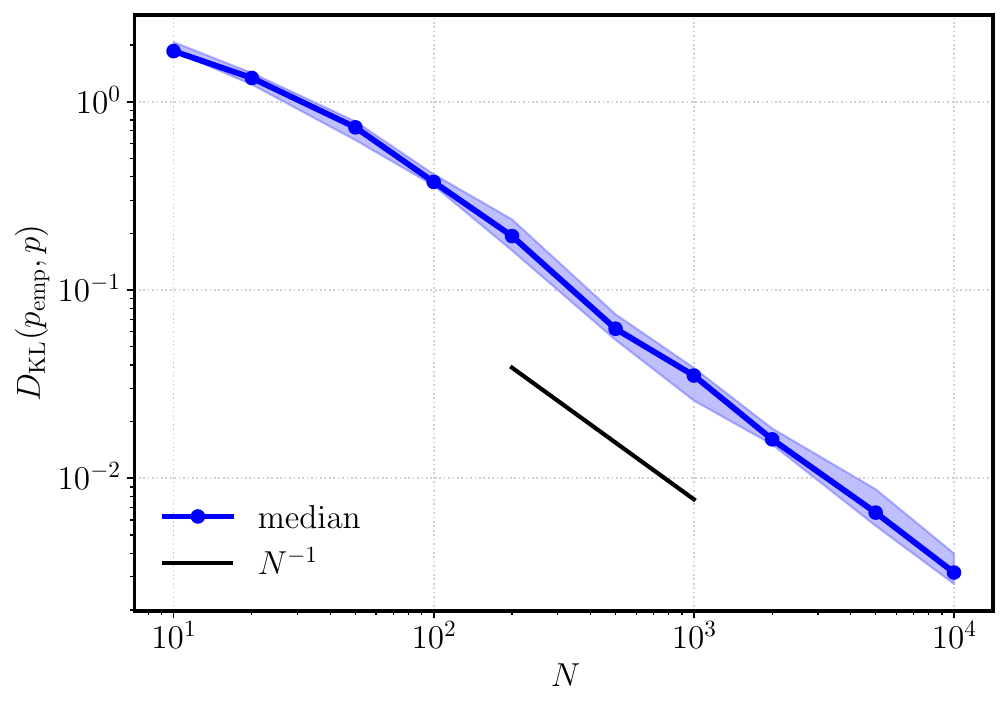}
\end{minipage}
\begin{minipage}{0.45\textwidth}
\centering
\includegraphics[scale=0.4]{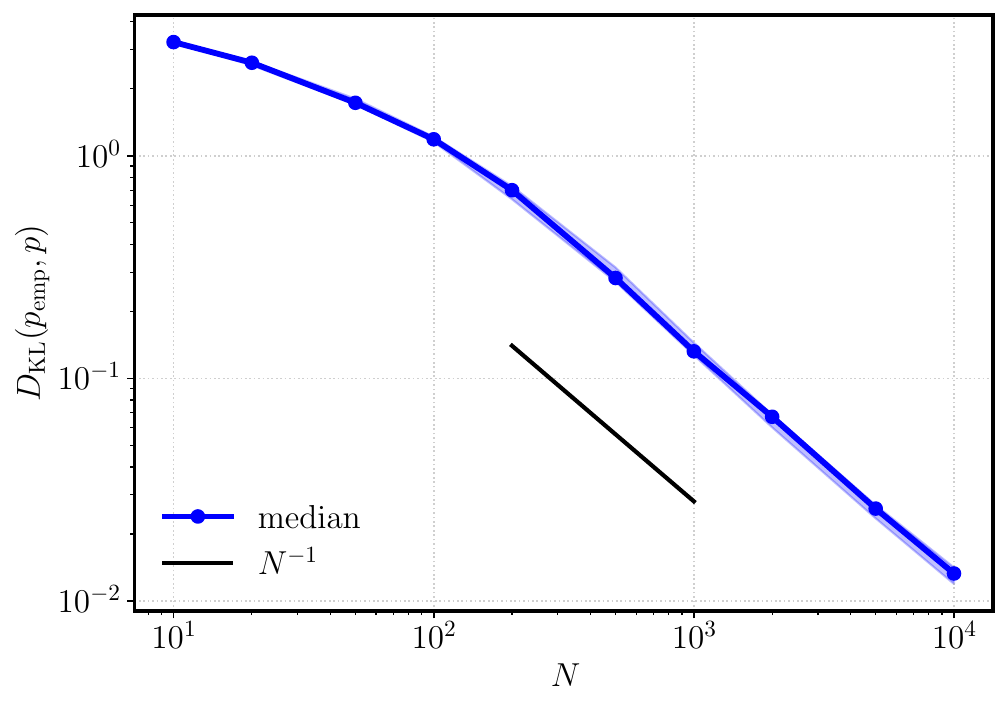}
\end{minipage}

\caption{Median KL divergence between the empirical and exact distributions as a function of $N$, the number of independently drawn samples. Shaded regions indicate the 16th–84th percentile range over independent Monte Carlo trials. The black line shows the expected $N^{-1}$ Monte Carlo convergence rate. }
\label{fig:GHZ_W_KL}
\end{figure}

We first tested the isoTNS sampling Algorithm~\ref{alg:iso_sample} by sequentially drawing independent samples to build up empirical distributions $p_{\mathrm{emp}}$ for the GHZ-state \eqref{eq:GHZ} and the W-state \eqref{eq:W}. We compared these empirical distributions with the analytically known distributions $p$ and measure their distance using KL-divergence 
\begin{equation} \label{eq:KL}
D_{\mathrm{KL}}(p_{\mathrm{emp}},p) = \sum_{\sigma \in \mathcal{S}} p_{\mathrm{emp}}(\sigma) \log\left(\frac{p_{\mathrm{emp}}(\sigma)}{p(\sigma)}\right), 
\end{equation} 
where $\mathcal{S}$ is the set of possible outcomes (2 for GHZ-state and $L^2$ for W-state). 
When contracting columns in step 3 of the sampling algorithm~\ref{alg:iso_sample} we use the zip-up algorithm with maximum bond-dimension 2, which yields no truncation error. With no truncation error, the isoTNS sampling algorithm is perfect \cite{Djuric2002} in the sense that each sample is independent and drawn directly from the underlying distribution. 

We considered both states on $8 \times 8$ and $16 \times 16$ lattices. For each state and system size we constructed $10$ empirical distributions using various numbers of samples $N$. The KL-divergence \eqref{eq:KL} as a function of $N$ is shown in Figure~\ref{fig:GHZ_W_KL}. The blue line indicate the median KL-divergence of the $10$ independent Monte Carlo trials and the shaded regions indicate the 16th–84th percentile range over independent trials. For the GHZ-state we find that the error of the empirical distribution is lower than that of the W-state when fixing the number of samples. This is because there are only two possible outcomes for GHZ state. For the W-state, the number of outcomes $L^2$ increases with the lattice size and hence the error increases with $L$ for a fixed number of samples. We observe that in all cases convergence follows the expected $N^{-1}$ Monte Carlo convergence rate. 

\begin{figure}[t]
\centering
\includegraphics[scale=0.4]{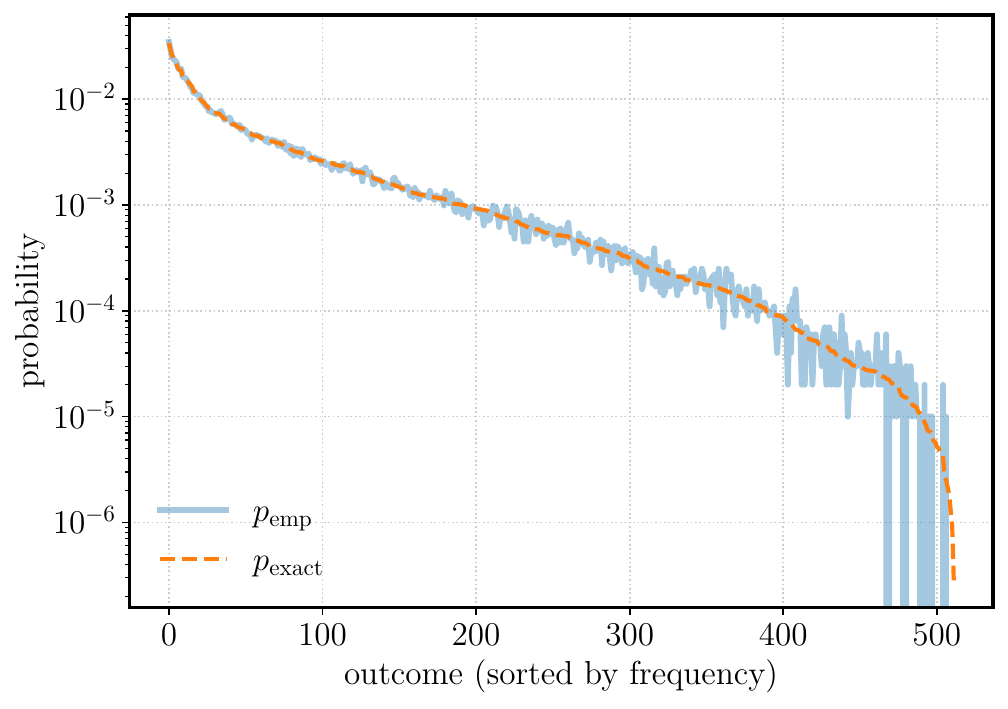}
\includegraphics[scale=0.4]{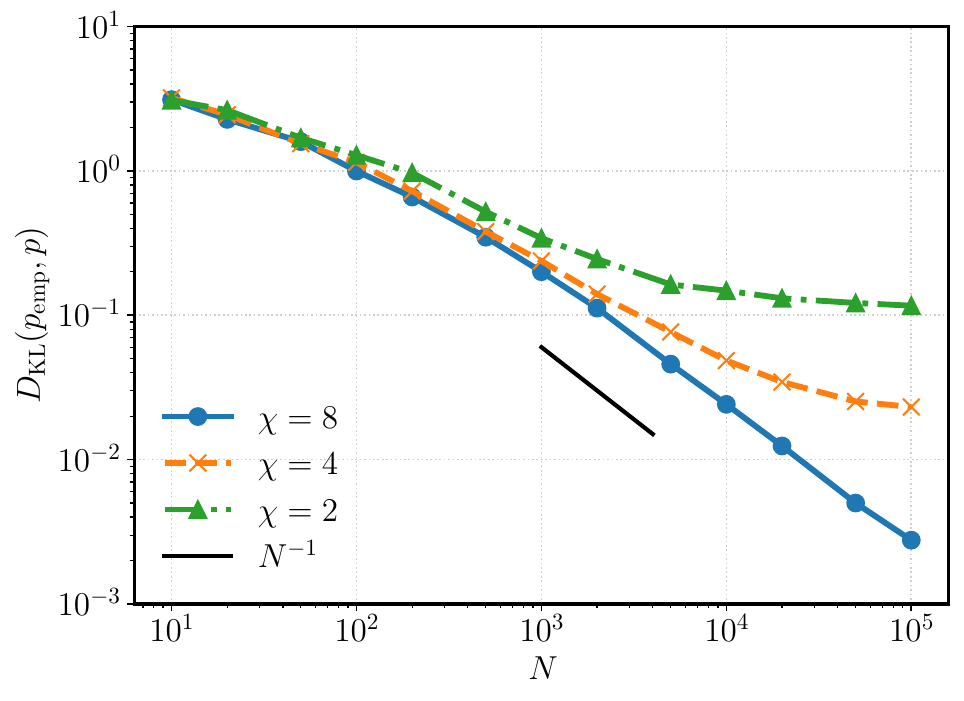}
\caption{
Left: Empirical distribution of random state from 10000 samples using the proposed isoTNS algorithm compared with the exact distribution. 
Right: KL divergence of the empirical distribution and the distribution computed from state-vector versus number of samples for a representative set of maximum bond-dimensions $\chi$ in the isoTNS sampling algorithm~\ref{alg:iso_sample}. } 
\label{fig:rand_error}
\end{figure}

Next we considered a randomly generated state on a $3 \times 3$ lattice. Since in this case the random distribution is not known a-priori analytically known, we limit our calculation to a small system size so that we may benchmark against a reference probability distribution computed using a state-vector formalism. 
We computed empirical distributions using bond-dimensions $\chi=8,4,2$ to study truncation error in sampling  when contracting rows with the zip-up method in Step 5 of Algorithm~\ref{alg:iso_sample}. Our results are shown Figure~\ref{fig:rand_error}. The left panel shows numerically exact probability distribution (orange dashed line) and the empirical distribution computed using isoTNS sampling with bond-dimension $\chi=8$ and $10^5$ samples. The right panel shows the convergence of empirical distributions as a function of the number of samples for each maximum bond-dimension $\chi=8,4,2$. For $\chi = 8$, there are no truncation errors in the isoTNS sampling algorithm and the convergence follows the $N^{-1}$ convergence rate of Monte Carlo. For bond-dimensions $\chi=4,2$ the convergence is limited by truncation errors in the isoTNS sampling algorithm. 

\subsection{Top-$K$ demonstration}

Next we \rev{validate} the top-$K$ sampling Algorithm~\ref{alg:iso_topK} with our suite of test states and their underlying probability distributions. 

\rev{
We began with the GHZ-state and W-state on lattices with side lengths $L=2$ up to $L=32$. For GHZ we ran the top-$K$ algorithm with $K=2$ and for W we used $K=L^2$. In all cases the top-$K$ algorithm recovered the correct distribution up to machine precision as expected. 
}


Next we applied the top-$K$ algorithm to an isoTNS generated by a random circuit on a lattice with side length $L=3$. As we have already seen in Section~\ref{sec:empirical_results} and Figure~\ref{fig:rand_error} the distribution has several high-probability states and many states with exponentially decaying probability as expected from a Porter-Thomas distribution \cite{Porter1956, Boixo2018, Arute2019}. We set $K=10$ to compute several high-probability states. We ran the top-$K$ algorithm three times with different maximum bond-dimensions $16, 12, 8$, which resulted in truncation errors (measured in the Frobenius norm) on row 2 in the top-$K$ algorithm of machine precision, $5.6\times 10^{-2}$, $1.5\times 10^{-1}$, respectively. 

\begin{figure}[t]
\centering
\includegraphics[scale=0.4]{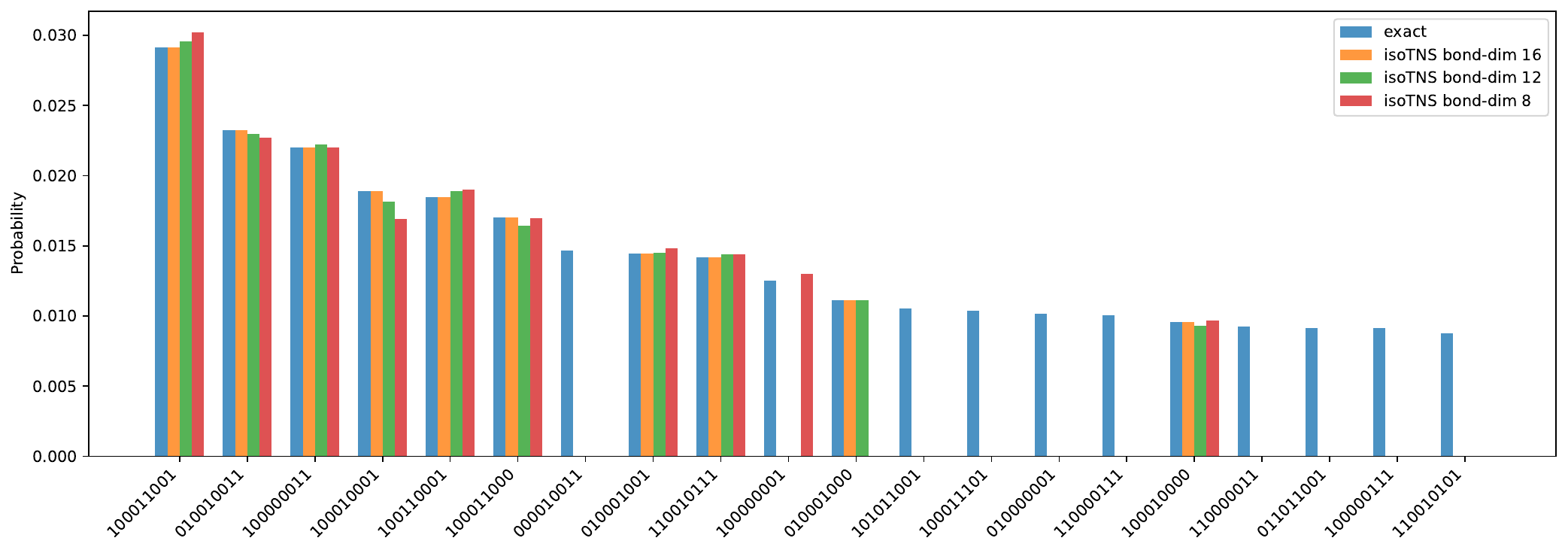}
\\
\includegraphics[scale=0.4]{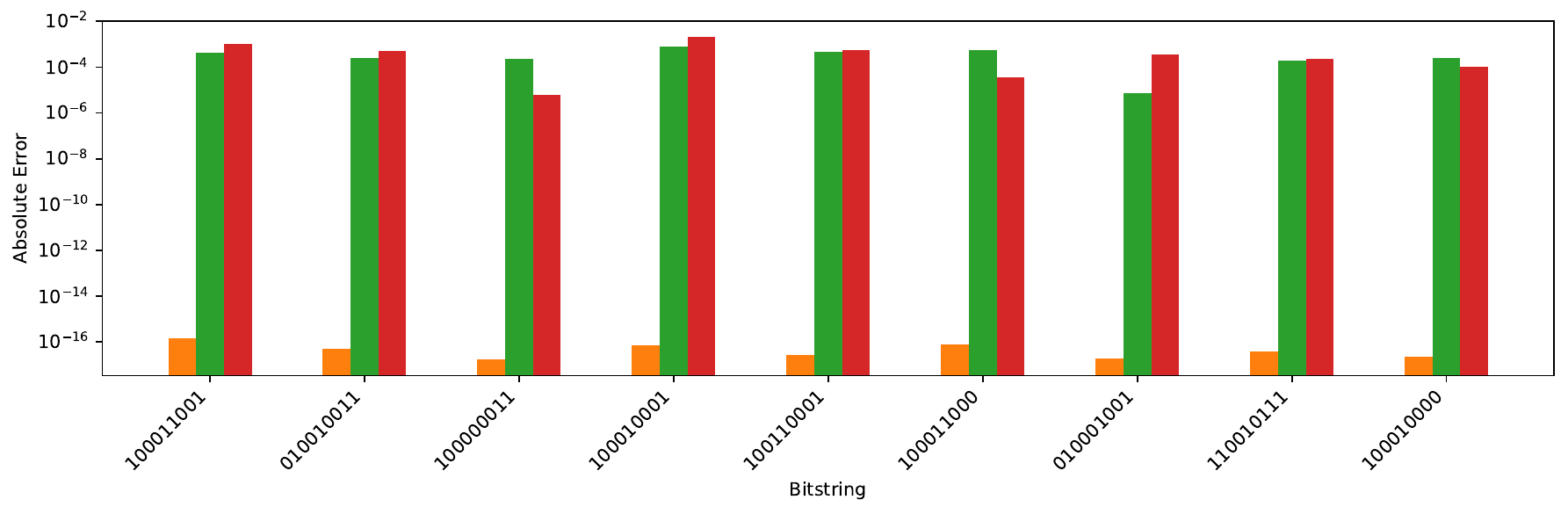}
\caption{\rev{Greedy top-$K$ isoTNS algorithm applied to a random state. With isoTNS bond dimension 16, 12, 8 the truncation error is machine precision, $5.6 \times 10^{-2}$, $1.5 \times 10^{-1}$, respectively. Top: Exact probabilities of the 20 highest-probability states compared with the states returned by the greedy top-$K$ algorithm. For $\chi=16$, the probabilities of the returned states agree with the exact values up to numerical precision. The remaining discrepancy is due to the greedy search not recovering the exact global top-$K$ set beyond the first 6 states. Bottom: Absolute differences between the exact and computed probabilities for the states returned by the top-$K$ algorithm.}
} 
\label{fig:topK_rand}
\end{figure}

The results are shown in Figure~\ref{fig:topK_rand}. The top panel compares the 20 highest-probability bitstrings of the exact distribution with the configurations returned by the greedy top-$K$ isoTNS algorithm. 
\rev{As discussed in \cref{sec:topK}, the top-$K$ procedure is greedy and is not guaranteed to recover the exact global top-$K$ states. For all bond dimensions considered, the algorithm correctly recovers the 6 most probable states. For $\chi=16$, the probabilities of the returned configurations agree with the exact probabilities up to numerical precision, so the remaining discrepancy in the top panel is not caused by truncation error. Rather, beyond the first six states, the greedy search misses some exact top-ranked configurations and instead returns other large probability configurations. For $\chi=12$ and $\chi=8$, there is an additional source of error from finite-bond-dimension truncation in the row contractions. This is quantified in the bottom panel, which shows the absolute differences between the exact and computed probabilities for the configurations returned by the top-$K$ algorithm. The $\chi=16$ errors are at numerical precision, whereas the $\chi=12$ and $\chi=8$ errors quantify algorithmic truncation errors.}


\begin{figure}[t]
\centering
\footnotesize{\hspace{.7cm} (a) \hspace{5.5cm} (b)}
\includegraphics[scale=0.4]{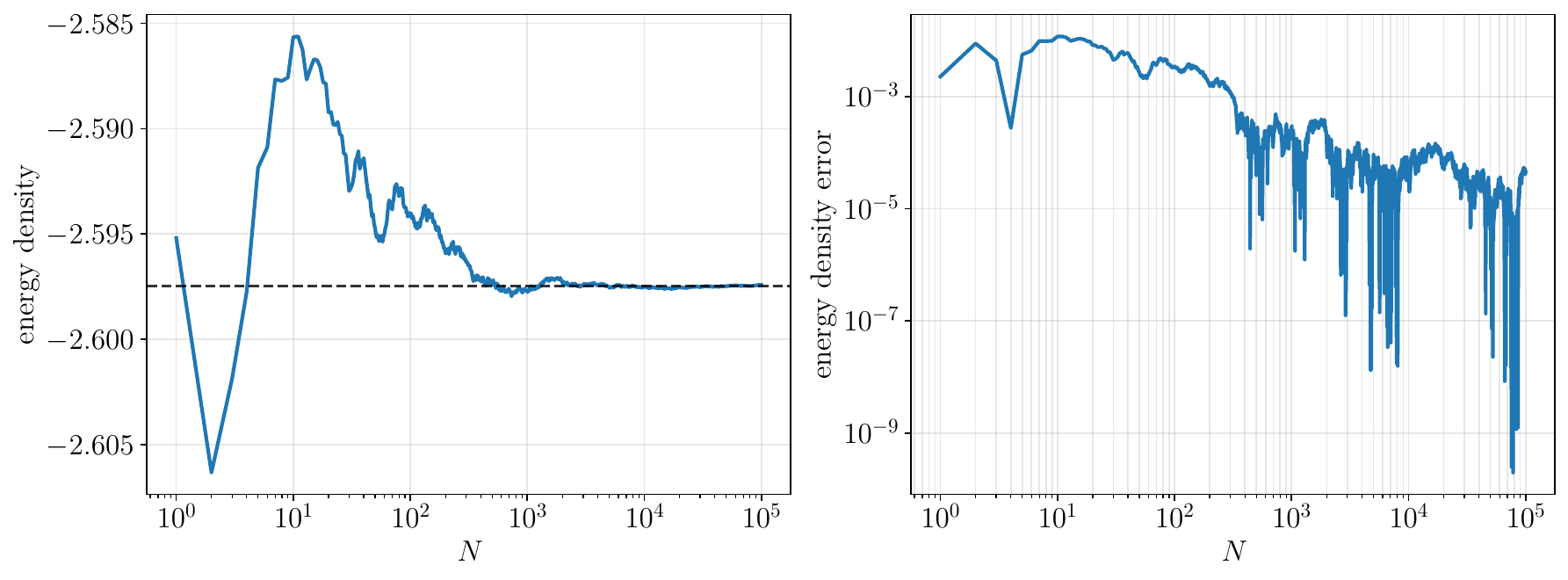} \\
\footnotesize{\hspace{.5cm} (c)} \\
\includegraphics[scale=0.4]{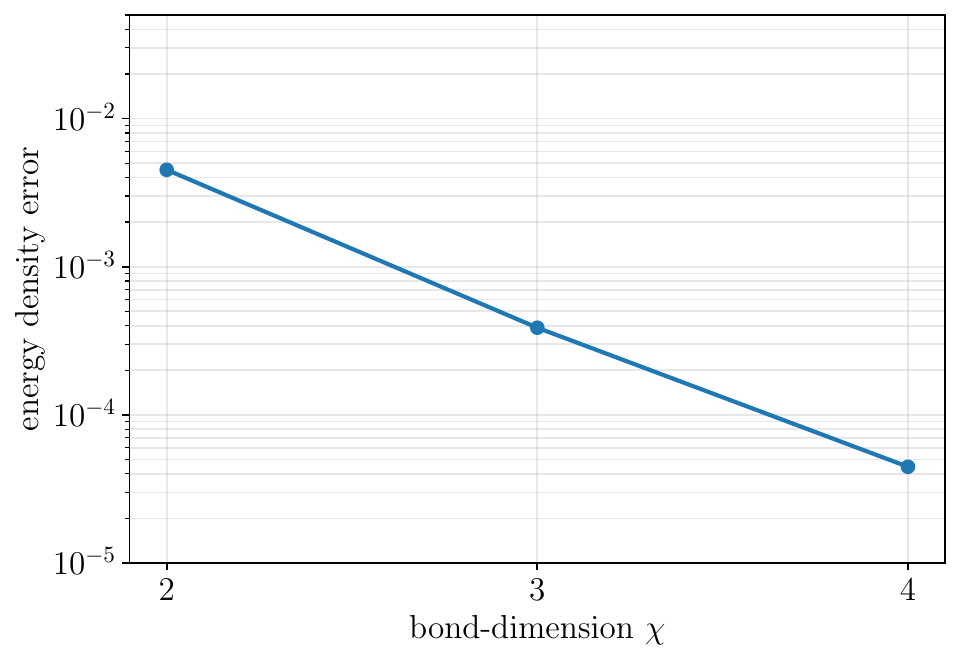}
\caption{
\rev{
METTS benchmark for the $3 \times 3$ transverse-field Ising model \eqref{eq:tfi} at $\beta=1.0$ and $J = 1.0, h = 2.5$. (a) Running estimate of the energy density from the isoTNS-METTS calculation as a function of the number of METTS samples for maximum bond dimensions $\chi=4$.
(b) Running absolute error of the energy density from the isoTNS-METTS calculation as a function of the number of METTS samples for maximum bond dimensions $\chi=4$. 
(c) Final absolute error in energy density versus maximum isoTNS bond-dimension $\chi$ using $10^5$ samples. 
}
} 
\label{fig:metts_tfim}
\end{figure}

\rev{
\subsection{METTS for the transverse-field Ising model}

Finally, let us test the effects of isometric tensor constraints and sampling in a realistic workflow with a more complex distribution than the analytic GHZ- and W- or random circuit distributions above. Specifically, we demonstrate the proposed sampling algorithm~\ref{alg:iso_sample} for computing thermal properties of the transverse-field Ising model 
\begin{equation} \label{eq:tfi}
H = -J \sum_{\langle i,j \rangle} \sigma_i^z \sigma_j^z + h \sum_i \sigma^x_i, 
\end{equation}
where $\sigma^z,\sigma^x$ are Pauli matrices, on a square lattice with open boundary conditions. As we mentioned in Section~\ref{sec:intro} such sampling procedures are a key sub-routine in the minimally entangled typical thermal states (METTS) algorithm~\cite{Stoudenmire2010}. We pair our proposed sampling algorithm with the TEBD2 algorithm, introduced in \cite{zaletel2020isometric}, to perform the imaginary time evolution required in METTS. We consider a $3 \times 3$ lattice where we can compare our results with benchmarks obtained from exact diagonalization. 

We take coupling constants $J = 1.0, h = 2.5$, inverse temperature $\beta = 1.0$, and generate a Markov chain of isoTNS METTS starting from a product state. This forms the approximate normalized typical state
\begin{equation} \label{eq:typical_state}
|\phi_\sigma\rangle \approx
\frac{e^{-\beta H/2}|\sigma\rangle}{\|e^{-\beta H/2}|\sigma\rangle\|},
\end{equation}
from which we measure the energy density in $|\phi_\sigma\rangle$, before collapsing $|\phi_\sigma\rangle$ to a new product state. The imaginary time evolution to obtain \eqref{eq:typical_state} is performed using the TEBD2 algorithm~\cite{zaletel2020isometric} with step-size $\delta \beta = 0.05$ and sequential Moses Move to shift the orthogonality center. The collapse step is performed with Algorithm~\ref{alg:iso_sample}. To examine algorithmic errors, we ran three calculations using different maximum bond-dimensions $\chi=2,3,4$. For each run, we discarded the first $10$ METTS for equilibration and use the subsequent states to estimate thermal energy density. 

To explicitly see how the algorithm converges, in Figure~\ref{fig:metts_tfim}(a), as a function of sample number together with the reference value, we display the running METTS estimate of the energy density from the calculation using maximum bond-dimension $\chi=4$. 
Figure~\ref{fig:metts_tfim}(b) then illustrates the running absolute error of the energy density from the isoTNS-METTS calculation as a function of the number of METTS samples for maximum bond dimensions $\chi = 4$. 
Finally, in Figure~\ref{fig:metts_tfim}(c), we show the final absolute error in energy density and as a function of the maximum isoTNS bond dimension $\chi$ for $10^5$ samples. For the parameters considered here, we find that the largest $\chi$ agrees with ED within statistical error, while smaller $\chi$ shows a systematic bias correlated with truncation error. 
}

\section{Conclusions} \label{sec:conclusions}
We introduced the first algorithms \rev{that leverage isometric constraints} to sample and compute high-probability configurations of distributions represented by 2D isometric tensor network states (isoTNS). We proposed two algorithms \rev{that} are non-trivial \rev{generalizations} of existing MPS sampling methods, e.g., Algorithm~\ref{alg:mps_sample}, to the isoTNS tensor network canonical form. The first algorithm performs independent sampling and yields a single configuration together with its associated probability. The second algorithm employs a greedy search strategy to identify $K$ high-probability configurations and their corresponding probabilities. In both cases we have bounded the algorithms' runtime as a polynomial in the system size, \rev{local Hilbert-space dimension,} virtual bond-dimensions, and $K$.


To validate and benchmark our results, we tested the proposed algorithms on a variety of examples with increasing complexity. In particular the GHZ- and W-states' probability distributions were calculated using both algorithms on up to 256 qubits. In addition to increasing system size, we explored a generically hard distributions which are generated by random unitary transformations. Our numerical experiments validate the scalability of the proposed algorithms and explored the effects of truncation, which are not present in MPS algorithms, introduced during the sampling procedures. 


\rev{
This work opens various new directions. While our algorithm straightforwardly generalizes to higher dimensional isometric forms, numerical validation and testing should first be performed on three-dimensional states~\cite{Tepaske2021}. We also expect further refinements in sampling strategies and in estimating probability distributions. For example, we did not study empirical distributions constructed from repeated applications of the top-$K$ algorithm, nor did we explore more intricate Monte-Carlo algorithms based on importance sampling schemes. In particular, it would be of interest to investigate whether Metropolis-type algorithms can improve efficiency and enable more rapid sampling of high-probability configurations.
} 

\rev{
More broadly, this work paves the way for incorporating Monte Carlo sampling into tensor network methods for quantum many-body systems. As an initial step in this direction, we demonstrated the use of the proposed sampling algorithm within a METTS calculation for the transverse-field Ising model. While this example illustrates the feasibility of combining isoTNS sampling with METTS, a systematic investigation of isoTNS-based METTS algorithms, including their accuracy, efficiency, and scalability, remains an important topic for future work. We anticipate that further developments along these lines will enable finite-temperature simulations, of area law quantum models, with isoTNS.
}

While our work did not succeed in collapsing the polynomial hierarchy~\cite{Bremner2011}, it does \rev{provide a tool that can} push the fundamental limits for classically sampling from what are considered hard distributions. For example, we expect our algorithms to play an important role in future verifications and extensions of quantum supremacy experiments~\cite{Arute2019}\rev{, especially for realistic experiments where correlations can be aggressively truncated commensurate with experimental noise~\cite{Zhou2020}}. It will therefore be of interest to further explore the fundamental limits of this class of sampling algorithms. Lastly, we expect our work to find applications in applied mathematics beyond many-body quantum physics. In particular, our approach is well suited for uncertainty quantification, in the same vein as tensor-train–based sampling algorithms~\cite{Dolgov2020,Chertkov2022,Batsheva2023,Novikov21}.

\section{Acknowledgments} \label{sec:Acknowledgments}

This work was supported by the U.S. Department of Energy, Office of Science, Office of Advanced Scientific Computing Research and Office of Basic Energy Sciences, Scientific Discovery through Advanced Computing (SciDAC) program under the CONNEQT project. A.D. and C.Y. also acknowledge support through the FASTMath Institute under U.S. Department of Energy Contract No. DE-AC02-05CH11231. 


\bibliographystyle{elsarticle-num}
\bibliography{refs}

@article{Djuric2002,
  title={Perfect sampling: a review and applications to signal processing},
  author={Djuric, Petar M and Huang, Yufei and Ghirmai, Tadesse},
  journal={IEEE Transactions on Signal processing},
  volume={50},
  number={2},
  pages={345--356},
  year={2002},
  publisher={IEEE}
}

@article{Verstraete2004,
  title={Renormalization algorithms for quantum-many body systems in two and higher dimensions},
  author={Verstraete, Frank and Cirac, J Ignacio},
  journal={arXiv preprint cond-mat/0407066},
  year={2004}
}

@article{Orus2014,
  title={A practical introduction to tensor networks: Matrix product states and projected entangled pair states},
  author={Or{\'u}s, Rom{\'a}n},
  journal={Ann. Phys.},
  volume={349},
  pages={117--158},
  year={2014},
  publisher={Elsevier}
}

@ARTICLE{Grafe2014,
  title     = "On-chip generation of high-order single-photon W-states",
  author    = "Gr{\"a}fe, Markus and Heilmann, Ren{\'e} and Perez-Leija,
               Armando and Keil, Robert and Dreisow, Felix and Heinrich,
               Matthias and Moya-Cessa, Hector and Nolte, Stefan and
               Christodoulides, Demetrios N and Szameit, Alexander",
  journal   = "Nat. Photonics",
  publisher = "Springer Science and Business Media LLC",
  volume    =  8,
  number    =  10,
  pages     = "791--795",
  month     =  oct,
  year      =  2014,
  language  = "en"
}

@article{Zhou2020,
  title = {What Limits the Simulation of Quantum Computers?},
  author = {Zhou, Yiqing and Stoudenmire, E. Miles and Waintal, Xavier},
  journal = {Phys. Rev. X},
  volume = {10},
  issue = {4},
  pages = {041038},
  numpages = {15},
  year = {2020},
  month = {Nov},
  publisher = {American Physical Society},
  doi = {10.1103/PhysRevX.10.041038},
  url = {https://link.aps.org/doi/10.1103/PhysRevX.10.041038}
}

@ARTICLE{Porter1956,
  title     = "Fluctuations of nuclear reaction widths",
  author    = "Porter, C E and Thomas, R G",
  journal   = "Phys. Rev.",
  publisher = "American Physical Society (APS)",
  volume    =  104,
  number    =  2,
  pages     = "483--491",
  month     =  oct,
  year      =  1956,
  copyright = "http://link.aps.org/licenses/aps-default-license"
}

@ARTICLE{Boixo2018,
  title     = "Characterizing quantum supremacy in near-term devices",
  author    = "Boixo, Sergio and Isakov, Sergei V and Smelyanskiy, Vadim N and
               Babbush, Ryan and Ding, Nan and Jiang, Zhang and Bremner,
               Michael J and Martinis, John M and Neven, Hartmut",
  journal   = "Nat. Phys.",
  publisher = "Springer Science and Business Media LLC",
  volume    =  14,
  number    =  6,
  pages     = "595--600",
  month     =  jun,
  year      =  2018,
  language  = "en"
}

@ARTICLE{Diker2025,
  title     = "Deterministic construction of arbitrary \textit{W} states with
               quadratically increasing number of two-qubit gates",
  author    = "Diker, F{\i}rat",
  journal   = "AIP Adv.",
  publisher = "AIP Publishing",
  volume    =  15,
  number    =  7,
  month     =  jul,
  year      =  2025,
  copyright = "https://creativecommons.org/licenses/by-nc-nd/4.0/",
  language  = "en"
}

@article{Lin2022,
  title = {Efficient simulation of dynamics in two-dimensional quantum spin systems with isometric tensor networks},
  author = {Lin, Sheng-Hsuan and Zaletel, Michael P. and Pollmann, Frank},
  journal = {Phys. Rev. B},
  volume = {106},
  issue = {24},
  pages = {245102},
  numpages = {23},
  year = {2022},
  month = {Dec},
  publisher = {American Physical Society}
}

@article{dai2025fermionic,
  title={Fermionic isometric tensor network states in two dimensions},
  author={Dai, Zhehao and Wu, Yantao and Wang, Taige and Zaletel, Michael P},
  journal={Phys. Rev. Lett.},
  volume={134},
  number={2},
  pages={026502},
  year={2025},
  publisher={APS}
}

@article{Perez-Garcia2007,
archivePrefix = {arXiv},
arxivId = {quant-ph/0608197},
author = {Perez-Garcia, D. and Verstraete, F. and Wolf, M. M. and Cirac, J. I.},
doi = {10.26421/qic7.5-6-1},
eprint = {0608197},
issn = {15337146},
journal = {Quantum Information and Computation},
month = {jul},
number = {5-6},
pages = {401--430},
primaryClass = {quant-ph},
title = {{Matrix product state representations}},
volume = {7},
year = {2007}
}

@article{Schollwock_2011,
   title={The density-matrix renormalization group in the age of matrix product states},
   volume={326},
   ISSN={0003-4916},
   url={http://dx.doi.org/10.1016/j.aop.2010.09.012},
   DOI={10.1016/j.aop.2010.09.012},
   number={1},
   journal={Annals of Physics},
   publisher={Elsevier BV},
   author={Schollwöck, Ulrich},
   year={2011},
   month=jan, pages={96–192} }

@ARTICLE{Bremner2011,
  title     = "Classical simulation of commuting quantum computations implies
               collapse of the polynomial hierarchy",
  author    = "Bremner, Michael J and Jozsa, Richard and Shepherd, Dan J",
  journal   = "Proc. Math. Phys. Eng. Sci.",
  publisher = "The Royal Society",
  volume    =  467,
  number    =  2126,
  pages     = "459--472",
  month     =  feb,
  year      =  2011,
  language  = "en"
}

@ARTICLE{Arute2019,
  title     = "Quantum supremacy using a programmable superconducting processor",
  author    = "Arute, Frank and others", 
  journal   = "Nature",
  publisher = "Springer Science and Business Media LLC",
  volume    =  574,
  number    =  7779,
  pages     = "505--510",
  month     =  oct,
  year      =  2019,
  language  = "en"
}

@article{zaletel2020isometric,
  title={Isometric tensor network states in two dimensions},
  author={Zaletel, Michael P and Pollmann, Frank},
  journal={Phys. Rev. Lett.},
  volume={124},
  number={3},
  pages={037201},
  year={2020},
  publisher={APS}
}

@misc{Hyatt2020,
      title={{DMRG} Approach to Optimizing Two-Dimensional Tensor Networks}, 
      author={Katharine Hyatt and E. M. Stoudenmire},
      year={2020},
      journal={arXiv:1908.08833}
}

@article{Haghshenas2019,
  title={Conversion of projected entangled pair states into a canonical form},
  author={Haghshenas, Reza and O'Rourke, Matthew J and Chan, Garnet Kin-Lic},
  journal={Phys. Rev. B},
  volume={100},
  number={5},
  pages={054404},
  year={2019},
  publisher={APS}
}

@article{Dektor2025,
  title={Computing excited states with isometric tensor networks in two-dimensions},
  author={Dektor, Alec and Chi, Runze and Van Beeumen, Roel and Yang, Chao},
  journal={arXiv:2502.19578},
  year={2025}
}

@article{Tepaske2021,
  title={Three-dimensional isometric tensor networks},
  author={Tepaske, Maurits SJ and Luitz, David J},
  journal={Phys. Rev. Research},
  volume={3},
  number={2},
  pages={023236},
  year={2021},
  publisher={APS}
}

@article{Wu2025,
  title={Alternating and Gaussian fermionic Isometric Tensor Network States},
  author={Wu, Yantao and Dai, Zhehao and Anand, Sajant and Lin, Sheng-Hsuan and Yang, Qi and Wang, Lei and Pollmann, Frank and Zaletel, Michael P},
  journal={arXiv:2502.10695},
  year={2025}
}

@article{Kadow2023,
  title={Isometric tensor network representations of two-dimensional thermal states},
  author={Kadow, Wilhelm and Pollmann, Frank and Knap, Michael},
  journal={Phys. Rev. B},
  volume={107},
  number={20},
  pages={205106},
  year={2023},
  publisher={APS}
}

@article{Soejima2020,
  title={Isometric tensor network representation of string-net liquids},
  author={Soejima, Tomohiro and Siva, Karthik and Bultinck, Nick and Chatterjee, Shubhayu and Pollmann, Frank and Zaletel, Michael P},
  journal={Phys. Rev. B},
  volume={101},
  number={8},
  pages={085117},
  year={2020},
  publisher={APS}
}

@article{Malz2025,
  title={Computational Complexity of Isometric Tensor-Network States},
  author={Malz, Daniel and Trivedi, Rahul},
  journal={PRX Quantum},
  volume={6},
  number={2},
  pages={020310},
  year={2025},
  publisher={APS}
}

@article{Sappler2025,
  title={Diagonal isometric form for tensor network states in two dimensions},
  author={Sappler, Benjamin and Kawano, Masataka and Zaletel, Michael P and Pollmann, Frank},
  journal={Physical Review B},
  volume={113},
  number={16},
  pages={165117},
  year={2026},
  publisher={APS}
}

@article{White2009,
  title={Minimally entangled typical quantum states at finite temperature},
  author={White, Steven R},
  journal={Physical review letters},
  volume={102},
  number={19},
  pages={190601},
  year={2009},
  publisher={APS}
}

@article{Stoudenmire2010,
  title={Minimally entangled typical thermal state algorithms},
  author={Stoudenmire, EM and White, Steven R},
  journal={New Journal of Physics},
  volume={12},
  number={5},
  pages={055026},
  year={2010},
  publisher={IOP Publishing}
}

@article{Wei2025,
  title={Numerical Optimization for Tensor Disentanglement},
  author={Wei, J. and Dektor, A. and Shen, C. and Wen, Z. and Yang, C.},
  journal={arXiv preprint arXiv:2508.19409},
  year={2025}
}

@article{Dumi,
  title = {Cloud Quantum Computing of an Atomic Nucleus},
  author = {Dumitrescu, E. F. and McCaskey, A. J. and Hagen, G. and Jansen, G. R. and Morris, T. D. and Papenbrock, T. and Pooser, R. C. and Dean, D. J. and Lougovski, P.},
  journal = {Phys. Rev. Lett.},
  volume = {120},
  issue = {21},
  pages = {210501},
  numpages = {6},
  year = {2018},
  month = {May},
  publisher = {American Physical Society},
  doi = {10.1103/PhysRevLett.120.210501}
}

@article{Lami2023,
  title={Quantum magic via perfect pauli sampling of matrix product states},
  author={Lami, Guglielmo and Collura, Mario},
  journal={arXiv preprint arXiv:2303.05536},
  year={2023}
}

@article{Ferris2012,
  title={Perfect sampling with unitary tensor networks},
  author={Ferris, Andrew J and Vidal, Guifre},
  journal={Physical Review B—Condensed Matter and Materials Physics},
  volume={85},
  number={16},
  pages={165146},
  year={2012},
  publisher={APS}
}

@article{Khaneja2001,
  title={Cartan decomposition of SU (2n) and control of spin systems},
  author={Khaneja, Navin and Glaser, Steffen J},
  journal={Chemical Physics},
  volume={267},
  number={1-3},
  pages={11--23},
  year={2001},
  publisher={Elsevier}
}

@article{Chertkov2022,
  title={Optimization of functions given in the tensor train format},
  author={Chertkov, Andrei and Ryzhakov, Gleb and Novikov, Georgii and Oseledets, Ivan},
  journal={arXiv preprint arXiv:2209.14808},
  year={2022}
}

@article{Batsheva2023,
  title={PROTES: probabilistic optimization with tensor sampling},
  author={Batsheva, Anastasiia and Chertkov, Andrei and Ryzhakov, Gleb and Oseledets, Ivan},
  journal={Advances in Neural Information Processing Systems},
  volume={36},
  pages={808--823},
  year={2023}
}

@InProceedings{Novikov21,
  title = 	 {Tensor-train density estimation},
  author =       {Novikov, Georgii S. and Panov, Maxim E. and Oseledets, Ivan V.},
  booktitle = 	 {Proceedings of the Thirty-Seventh Conference on Uncertainty in Artificial Intelligence},
  pages = 	 {1321--1331},
  year = 	 {2021},
  editor = 	 {de Campos, Cassio and Maathuis, Marloes H.},
  volume = 	 {161},
  series = 	 {Proceedings of Machine Learning Research},
  month = 	 {27--30 Jul},
  publisher =    {PMLR}
}

@article{Dolgov2020,
  title={Approximation and sampling of multivariate probability distributions in the tensor train decomposition},
  author={Dolgov, Sergey and Anaya-Izquierdo, Karim and Fox, Colin and Scheichl, Robert},
  journal={Statistics and Computing},
  volume={30},
  number={3},
  pages={603--625},
  year={2020},
  publisher={Springer}
}

@article{Vieijra2021,
  title={Direct sampling of projected entangled-pair states},
  author={Vieijra, Tom and Haegeman, Jutho and Verstraete, Frank and Vanderstraeten, Laurens},
  journal={Physical Review B},
  volume={104},
  number={23},
  pages={235141},
  year={2021},
  publisher={APS}
}

@article{Gray2018,
    title={quimb: a python library for quantum information and many-body calculations},
    author={Gray, Johnnie},
    journal={Journal of Open Source Software},
    year = {2018},
    volume={3}, number={29}, pages={819},
    doi={10.21105/joss.00819},
}

@misc{isoTNSCode,
  author       = {A. Dektor},
  title        = {iso{TNS}\_sampling},
  year         = {2026},
  howpublished = {\url{https://github.com/adektor/isoTNS\_sampling}},
  note         = {Accessed: 2026-02-01}
}

@article{Jalowiecki2021,
  title={Brute-forcing spin-glass problems with CUDA},
  author={Ja{\l}owiecki, Konrad and Rams, Marek M and Gardas, Bart{\l}omiej},
  journal={Computer Physics Communications},
  volume={260},
  pages={107728},
  year={2021},
  publisher={Elsevier}
}

@article{Rams2021,
  title={Approximate optimization, sampling, and spin-glass droplet discovery with tensor networks},
  author={Rams, Marek M and Mohseni, Masoud and Eppens, Daniel and Ja{\l}owiecki, Konrad and Gardas, Bart{\l}omiej},
  journal={Physical Review E},
  volume={104},
  number={2},
  pages={025308},
  year={2021},
  publisher={APS}
}

@article{Vieijra2022,
  title={Generative modeling with projected entangled-pair states},
  author={Vieijra, Tom and Vanderstraeten, Laurens and Verstraete, Frank},
  journal={arXiv preprint arXiv:2202.08177},
  year={2022}
}

@article{Causer2023,
  title={Optimal sampling of dynamical large deviations in two dimensions via tensor networks},
  author={Causer, Luke and Ba{\~n}uls, Mari Carmen and Garrahan, Juan P},
  journal={Physical Review Letters},
  volume={130},
  number={14},
  pages={147401},
  year={2023},
  publisher={APS}
}

@article{Rudolph2025,
  title={Simulating and sampling from quantum circuits with 2D tensor networks},
  author={Rudolph, Manuel S and Tindall, Joseph},
  journal={arXiv preprint arXiv:2507.11424},
  year={2025}
}

\appendix

\rev{
\section{isoTNS construction of GHZ and W-states} \label{sec:isoTNS_construction}

To prepare the GHZ state \eqref{eq:GHZ} we first construct a PEPS representation \eqref{eq:peps} by performing a linear combination of the all zero state and all one state to obtain a PEPS with all bond-dimensions equal to 2. Then we transform this PEPS into an isoTNS \eqref{eq:2d_iso_tns} with orthogonality center in the upper left corner of the lattice using the sequential Moses Move~\cite{zaletel2020isometric}. We use a sufficiently large bond-dimension to not introduce errors during the sequential Moses Move. The maximum dimension of each bond in the resulting isoTNS representation is $2$. 

Constructing the W-state following the same steps we took for the GHZ-state involves constructing a PEPS as a linear combination of $L^2$ product states, resulting in excessively large bond-dimensions. Applying the Moses Move to such a PEPS is expensive and can introduce errors into the isoTNS representation of the state. Instead, we prepare the W-state as an isoTNS using quantum circuits that parameterize the single excitation subspace. Practically, as we describe below, a hyperspherical parameterization is realized with a compact amplitude-shifting circuit. Begin by exciting the orthogonality center's physical spin. We then wish to shift amplitude from one spin to the next, that is $\ket{1,0} \rightarrow \alpha \ket{1,0} + \beta \ket{0,1}$. This is performed first by a controlled-$Y$ rotation followed by a CNOT gate, which conditionally removes double excitations~\cite{Grafe2014, Dumi, Diker2025}. That is, first set $\alpha = \cos{\theta/2}$ by acting with $C_1 R_{Y_2}(\theta)\ket{1,0} = \cos{\theta/2} \ket{1,0} + \sin{\theta/2} \ket{1,1}$, where $C_i$ denotes a gate conditioned on spin $i$. Following this, the reversed CNOT ($C_2 X_1$) results in $\beta=\sin{\theta/2}$. That is, $C_2 R_{X_1} (\cos{\theta/2} \ket{1,0} + \sin{\theta/2} \ket{1,1}) = \cos{\theta/2} \ket{1,0} + \sin{\theta/2} \ket{0,1}$. To generate the W-state, we perform this circuit in serial so that, except for the first and final sites, each spin is the target once and the control once. To get a uniform distribution, the controlled-Y rotation angles are set as $\theta_i = 2 \arccos(\sqrt{1/(L^2-i)})$ where $0\leq i \leq L^2-1$ iteratively increases by $1$. Other angles could be chosen to describe generalized single-particle states (e.g., free fermions) but for benchmarking purposes, the W-state suffices. 
}

\end{document}